\documentclass[aps,prc,twocolumn,groupedaddress,floatfix,superscriptaddress,showpacs]{revtex4} 

\usepackage{graphicx}
\usepackage{amsmath}
\usepackage{lineno}

\usepackage{epsfig}

\usepackage{fancyhdr}
\usepackage{pslatex}
\usepackage{color}
\usepackage{amsmath, amsthm, amssymb}
\usepackage{natbib}
\usepackage{multirow}

\begin{document}



\title{Beam energy dependent two-pion interferometry and the freeze-out eccentricity of pions in heavy ion collisions at STAR}



\affiliation{AGH University of Science and Technology, Cracow, Poland}
\affiliation{Argonne National Laboratory, Argonne, Illinois 60439, USA}
\affiliation{University of Birmingham, Birmingham, United Kingdom}
\affiliation{Brookhaven National Laboratory, Upton, New York 11973, USA}
\affiliation{University of California, Berkeley, California 94720, USA}
\affiliation{University of California, Davis, California 95616, USA}
\affiliation{University of California, Los Angeles, California 90095, USA}
\affiliation{Universidade Estadual de Campinas, Sao Paulo, Brazil}
\affiliation{Central China Normal University (HZNU), Wuhan 430079, China}
\affiliation{University of Illinois at Chicago, Chicago, Illinois 60607, USA}
\affiliation{Cracow University of Technology, Cracow, Poland}
\affiliation{Creighton University, Omaha, Nebraska 68178, USA}
\affiliation{Czech Technical University in Prague, FNSPE, Prague, 115 19, Czech Republic}
\affiliation{Nuclear Physics Institute AS CR, 250 68 \v{R}e\v{z}/Prague, Czech Republic}
\affiliation{Frankfurt Institute for Advanced Studies FIAS, Germany}
\affiliation{Institute of Physics, Bhubaneswar 751005, India}
\affiliation{Indian Institute of Technology, Mumbai, India}
\affiliation{Indiana University, Bloomington, Indiana 47408, USA}
\affiliation{Alikhanov Institute for Theoretical and Experimental Physics, Moscow, Russia}
\affiliation{University of Jammu, Jammu 180001, India}
\affiliation{Joint Institute for Nuclear Research, Dubna, 141 980, Russia}
\affiliation{Kent State University, Kent, Ohio 44242, USA}
\affiliation{University of Kentucky, Lexington, Kentucky, 40506-0055, USA}
\affiliation{Korea Institute of Science and Technology Information, Daejeon, Korea}
\affiliation{Institute of Modern Physics, Lanzhou, China}
\affiliation{Lawrence Berkeley National Laboratory, Berkeley, California 94720, USA}
\affiliation{Massachusetts Institute of Technology, Cambridge, Massachusetts 02139-4307, USA}
\affiliation{Max-Planck-Institut f\"ur Physik, Munich, Germany}
\affiliation{Michigan State University, East Lansing, Michigan 48824, USA}
\affiliation{Moscow Engineering Physics Institute, Moscow Russia}
\affiliation{National Institute of Science Education and Research, Bhubaneswar 751005, India}
\affiliation{Ohio State University, Columbus, Ohio 43210, USA}
\affiliation{Old Dominion University, Norfolk, Virginia 23529, USA}
\affiliation{Institute of Nuclear Physics PAN, Cracow, Poland}
\affiliation{Panjab University, Chandigarh 160014, India}
\affiliation{Pennsylvania State University, University Park, Pennsylvania 16802, USA}
\affiliation{Institute of High Energy Physics, Protvino, Russia}
\affiliation{Purdue University, West Lafayette, Indiana 47907, USA}
\affiliation{Pusan National University, Pusan, Republic of Korea}
\affiliation{University of Rajasthan, Jaipur 302004, India}
\affiliation{Rice University, Houston, Texas 77251, USA}
\affiliation{University of Science and Technology of China, Hefei 230026, China}
\affiliation{Shandong University, Jinan, Shandong 250100, China}
\affiliation{Shanghai Institute of Applied Physics, Shanghai 201800, China}
\affiliation{SUBATECH, Nantes, France}
\affiliation{Temple University, Philadelphia, Pennsylvania 19122, USA}
\affiliation{Texas A\&M University, College Station, Texas 77843, USA}
\affiliation{University of Texas, Austin, Texas 78712, USA}
\affiliation{University of Houston, Houston, Texas 77204, USA}
\affiliation{Tsinghua University, Beijing 100084, China}
\affiliation{United States Naval Academy, Annapolis, Maryland, 21402, USA}
\affiliation{Valparaiso University, Valparaiso, Indiana 46383, USA}
\affiliation{Variable Energy Cyclotron Centre, Kolkata 700064, India}
\affiliation{Warsaw University of Technology, Warsaw, Poland}
\affiliation{University of Washington, Seattle, Washington 98195, USA}
\affiliation{Wayne State University, Detroit, Michigan 48201, USA}
\affiliation{Yale University, New Haven, Connecticut 06520, USA}
\affiliation{University of Zagreb, Zagreb, HR-10002, Croatia}

\author{L.~Adamczyk}\affiliation{AGH University of Science and Technology, Cracow, Poland}
\author{J.~K.~Adkins}\affiliation{University of Kentucky, Lexington, Kentucky, 40506-0055, USA}
\author{G.~Agakishiev}\affiliation{Joint Institute for Nuclear Research, Dubna, 141 980, Russia}
\author{M.~M.~Aggarwal}\affiliation{Panjab University, Chandigarh 160014, India}
\author{Z.~Ahammed}\affiliation{Variable Energy Cyclotron Centre, Kolkata 700064, India}
\author{I.~Alekseev}\affiliation{Alikhanov Institute for Theoretical and Experimental Physics, Moscow, Russia}
\author{J.~Alford}\affiliation{Kent State University, Kent, Ohio 44242, USA}
\author{C.~D.~Anson}\affiliation{Ohio State University, Columbus, Ohio 43210, USA}
\author{A.~Aparin}\affiliation{Joint Institute for Nuclear Research, Dubna, 141 980, Russia}
\author{D.~Arkhipkin}\affiliation{Brookhaven National Laboratory, Upton, New York 11973, USA}
% ...

\collaboration{STAR Collaboration}\noaffiliation




\date{\today}

\begin{abstract}
  We present results of analyses of two-pion interferometry in Au+Au collisions
at $\sqrt{s_{NN}}$ = 7.7, 11.5, 19.6, 27, 39, 62.4 and 200 GeV measured in the 
STAR detector as part of the RHIC Beam Energy Scan program.  The
extracted correlation lengths (HBT radii) are studied as a function of beam energy,
azimuthal angle relative to the reaction plane, centrality, and transverse mass
($m_{T}$) of the particles.  The azimuthal analysis allows 
extraction of the eccentricity of the entire fireball at kinetic freeze-out.  
The energy dependence of this observable is expected to be sensitive to changes in the equation of state.
A new global fit method is studied as an alternate method to directly measure the 
parameters in the azimuthal analysis.  The eccentricity shows a monotonic 
decrease with beam energy that is qualitatively consistent with the trend from 
all model predictions and quantitatively consistent with a hadronic transport model.
\end{abstract}

\pacs{25.75.Gz, 25.75.Nq}

\maketitle

\section{Introduction}\label{S1}


  The Beam Energy Scan program performed at the Relativistic Heavy Ion
Collider (RHIC) in 2010 and 2011 was designed to map features expected to 
appear in the QCD phase diagram \cite{BESdoc}.  At the highest RHIC energies 
evidence suggests that the matter
formed in heavy ion collisions is a hot, strongly 
coupled fluid of deconfined quarks and gluons (sQGP)
\cite{QGPSTAR,QGPPHENIX,QGPPHOBOS,QGPBRAHMS}, with rather low
chemical potential, $\mu_{B}$.  The nature of this phase transition is likely
a smooth, rapid cross-over transition
\cite{CROSSOVER,FlowPapersCrossoverA,WBcrossoverA,HotQCDcrossoverA}.  As the beam energy 
is lowered, the matter produced near mid-rapidity evolves through regions of 
the phase diagram at larger $\mu_{B}$.  At higher chemical potentials
there are predictions from lattice calculations of a change to a first-order
phase transition with an associated latent heat 
\cite{1stOrderTransitionA,
1stOrderTransitionD,1stOrderTransitionE,1stOrderTransitionF,
1stOrderTransitionG,1stOrderTransitionH,1stOrderTransitionI}
and a critical point at some intermediate chemical potential \cite{CPatHighMuB}.  
The relative amounts 
of time the matter spends in an sQGP, mixed or hadronic phase may imprint a 
signal on observables that are sensitive to the equation of state
\cite{EosImprintsOnObservables}.  It is important, therefore, to study such 
observables as a function of beam energy both to search for possible 
non-monotonic behavior (which could indicate interesting physical changes in some 
aspect of the collisions) and to provide more stringent experimental guidance to 
theory and models.  The sizes and shapes that describe the matter produced 
in the collisions at freeze-out provide just this type of observable 
\cite{ShapeAnalysis}.

Results of two-pion interferometry analyses (often referred to as HBT analyses)
are presented in this paper
as a function of beam energy.  Hanbury Brown and Twiss invented 
the intensity interferometry technique to measure sizes of nearby stars \cite{HBTstellarRef1}.
The technique was extended to particle physics \cite{GGLPppbarRef1} to  
study angular distributions of pion pairs in $p\bar{p}$ annihilations, finding
that quantum statistics caused 
an enhancement in pairs with low relative momentum. 
In subsequent HBT analyses the method 
has evolved into a precision tool for measuring space-time properties of the 
regions of homogeneity at kinetic freeze-out in heavy ion collisions \cite{AnnRevHbt}.  
 Two-pion interferometry yields HBT radii that describe the geometry of these
regions of homogeneity (regions that emit correlated pion pairs).
The observation that HBT radii increase for more central collisions is
attributed to the increasing volume of the source, an example of how HBT can 
probe spatial sizes and shapes \cite{AnnRevHbt}.  In addition to the spatial 
shape and size of these regions from which particle pairs are emitted, 
space-momentum correlations induced by collective (and anisotropic) flow 
\cite{LisaRetiere} may imprint patterns on the results.  For instance, the HBT 
radii exhibit a systematic decrease with mean pair transverse momentum, $k_{T}$, 
which has been attributed to transverse and longitudinal flow
\cite{LisaRetiere,CorrespondenceHBT}.  The presence of flow induces 
space-momentum correlations so that the size of the regions emitting particles 
does not correspond to the entire fireball created in a collision \cite{AnnRevHbt,LisaRetiere,
CorrespondenceHBT}.  In standard HBT analyses, integrated over azimuthal angle 
relative to the reaction plane, the extracted source sizes correspond only to 
some smaller region of the total volume; the higher the 
transverse momentum, $k_{T}$, the smaller the radii describing the volume 
emitting the particles \cite{AnnRevHbt}.  However, in HBT analyses performed relative to the 
reaction plane, sinusoidal variations in the shape of these smaller source 
regions can be connected to the overall shape of the entire fireball 
\cite{LisaRetiere,CorrespondenceHBT}.

  Previous HBT analyses from various experiments have led to a large world 
data set for standard, non-azimuthal HBT results 
at AGS \cite{RvsRootSE802,RvsRootSE895,RvsRootSE866} and SPS \cite{RvsRootSCERES,RvsRootSNA44,RvsRootSNA49,RvsRootSWA98},
as well as top RHIC energies \cite{RvsRootSPHOBOS,PRCstarAuAu200,CuCuAuAu,ppFemto}, and 
at the LHC \cite{AliceHbtPLB2010,AliceHbtAdamsJPhysG2011,AliceHbtGramlingAIPConfProc2012}.  
In contrast, only a few azimuthal 
HBT results have been reported previously by E895 \cite{E895aziHbtData}, 
STAR \cite{PRLstarAziAuAu200} and CERES \cite{CERESaziHbtPRC}.  While 
the results suggested possible non-monotonic behavior in the freeze-out shape 
of the collisions with a minimum appearing around a collision energy per nucleon
of 17.3 GeV, the sparse amount 
of data coming from several different experiments could not allow one to draw a 
definite conclusion \cite{CERESaziHbtPRC}.  In this paper, the results of 
azimuthally integrated HBT analyses are placed in the context of the world 
data set reproducing the low energy and high energy results and filling in the 
intermediate energy region with results from a single detector and identical 
analysis techniques.  The azimuthally differential HBT results are also 
presented across this wide range of energies allowing extraction of the beam 
energy dependence of the transverse eccentricity at freeze-out.  



  In the case of the azimuthally differential analysis, a new global fit method
is developed.  The technique, described in this paper, uses a Gaussian
parameterization.  However, several correlation functions constructed in
azimuthal bins relative to the reaction plane are fit simultaneously.  This
allows direct extraction of Fourier coefficients that describe the observed
sinusoidal variations in the shape of the regions of homogeneity that emit pion
pairs.  This technique avoids correlated errors that arise from a correction 
for finite-bin-width and event plane resolution effects and it is more
robust in some cases where statistics and event plane resolutions are low.  The
global fit method provides the most reliable estimate of the shape of the
fireball at kinetic freeze-out which, as described in the next section, is used
to search for a change in the type of phase transition at lower energies.
The experimental results of this study are presented in Sec.~\ref{S5b3}.

  
\section{Collision evolution and freeze-out shape}\label{S2}

  A primary theme explored in this analysis is the connection between the
type of phase transition the system experiences and the shape 
of the collision during kinetic freeze-out. 
Therefore, in this section we explore the relationship between the underlying 
physics and the final shape achieved in the collisions.  
In non-central collisions, the second order anisotropy of the participant zone
(in the transverse plane) is an ellipse extended out of the reaction plane
(the plane containing the impact parameter and beam direction).  Initial state 
fluctuations in positions of participant nucleons may cause deviations from a 
precise elliptical shape \cite{InitialShapeFluctuations}.  Nevertheless, the 
initial shape is approximately elliptical and can be estimated using Monte 
Carlo Glauber calculations.
Due to the anisotropic shape and the speed of sound, $c^{2}_{s}=\partial
p/\partial e$ (where p is pressure and e is energy density), larger initial
pressure gradients appear along the short axis.
These stronger in-plane pressure gradients 
drive preferential in-plane expansion, thereby reducing the eccentricity.
The system must evolve to a less out-of-plane extended freeze-out shape.
Longer lifetimes, stronger pressure gradients, or both, would lead to expansion 
to an even more round
or even in-plane extended (negative eccentricity) shape at kinetic freeze-out.  
It would be expected that increasing the beam energy would lead to longer
lifetimes and pressure gradients and so a monotonically decreasing excitation
function for the freeze-out eccentricity would be expected \cite{ShapeAnalysis}.  
In fact, all transport and hydrodynamic models predict a monotonic decrease in 
the energy ranges studied here.  

  There is, however, another consideration related to the equation of state.
If the nature of the phase transition changes from a smooth cross-over at high 
energy to a first-order transition at lower energy, the matter will evolve 
through a mixed-phase regime (associated with a latent heat) during which the 
pressure gradients vanish ($c^{2}_{s}=0$).  Outside of a mixed-phase regime,
the equation of state 
has even stronger pressure gradients ($c^{2}_{s}=1/3$) in the sQGP phase than 
the hadronic phase ($c^{2}_{s}=1/6$) \cite{EoSsoundspeed,KolbHeinzHydro}.  
As the collision energy is varied, the collisions evolve along different
trajectories through the $T$-$\mu_{B}$ phase diagram.  At low energy the system
may evolve through a first-order phase transition and the length of time spent 
in the various phases may alter the amount of expansion that takes place prior 
to freeze-out \cite{KolbHeinzHydro}.  It is possible that a non-monotonic freeze-out 
shape might be observed as a result.  In fact, it was speculated in 
\cite{ShapeAnalysis} that the possible minimum observed in the previously
available 
freeze-out eccentricity measurements might be caused by entrance into a mixed-phase regime around a 
minimum, followed by a maximum at higher energy above which the system achieves 
complete deconfinement (and the strong pressure gradients reappear).  Measuring 
the energy dependence of the freeze-out shape therefore allows one to probe 
interesting physics related to both the equation of state and dynamical 
processes that drive the evolution of the collisions.

\section{Experimental setup and event, track, and pair selections}\label{S3}

\subsection{STAR detector}\label{S3a}

  The STAR detector \cite{STARdetector} was used to reconstruct Au+Au collisions 
provided at $\sqrt{s_{NN}}$ = 7.7, 11.5, 19.6, 27, 39, 62.4 and 200 GeV as part 
of a first phase of the Beam Energy Scan program.  The main detector used in 
this analysis is the Time Projection Chamber (TPC) 
\cite{STARTPCNIMA,STARTPCNuclPhysA}, which allows reconstruction of the momentum 
of charged particles used for event plane determination, including the charged 
pions used in the HBT analyses.  The TPC 
covers the pseudorapidity range $|\eta|<1$
and has full 2$\pi$ azimuthal acceptance. It is located 
inside a 0.5 T solenoidal magnetic field for all energies to aid in identifying 
the charge, momentum, and species of each track.  Zero Degree Calorimeters, 
Beam-Beam Counters and/or Vertex Position Detectors, located at large rapidities near the 
beam line, were tuned online to collect high statistics, minimum bias data sets 
at each energy.  Measuring coincidences of spectator particles in the subsystems
allows selection of collisions that occur near the center of the detector.

\begin{table}
\begin{center}
\begin{tabular}{lll}
\hline \hline
\\[3pt]
$\sqrt{s_{NN}}$ (GeV)      &     $|V_{Z}|$ (cm)      &     $N_{\mathrm{events}}$ ($10^{6}$) \\[3pt]
\hline
\\[3pt]
 7.7      &   $<70$       &     3.9 \\
11.5      &   $<50$       &    10.7 \\
19.6      &   $<30$       &    15.4 \\
27        &   $<30$       &    30.8 \\
39        &   $<30$       &     8.8 \\
62.4      &   $<30$       &    10.1 \\
200       &   $<30$       &    11.6 \\[3pt]
\hline \hline
\end{tabular}
\caption{\label{T1} Number of analyzed events and $z$-vertex range, $V_{Z}$, at each energy.}
\end{center}
\end{table}

\subsection{Event selection}\label{S3b}

  Events included in the analysis were selected using the reconstructed 
vertex position.  The radial vertex position ($V_{R}=\sqrt{V^{2}_{X}+V^{2}_{Y}}$) 
was required to be less than 2 cm to reject collisions with the beam pipe.  The 
vertex position along the beam direction, $V_{Z}$, was required to be near the center of 
the detector as summarized in Table~\ref{T1}, with larger ranges at 7.7 and 11.5 
GeV to maximize statistics.  The number of events at each energy used in
this analysis are also listed in Table~\ref{T1}.


  The events were binned in different centrality ranges based on multiplicity
as described in \cite{StarBesFlowEccentricity}.  For the azimuthal HBT analysis, data in the 0-5\%,
5-10\%, 10-20\%, 20-30\%, and 30-40\% centrality bins were used.  For the
non-azimuthal HBT analysis, additional 40-50\%, 50-60\% and 60-70\% bins were
also studied.

\subsection{Particle selection}\label{S3c}

  Tracks were selected in three rapidity ranges: $-1<y<-0.5$ (backward 
rapidity), $-0.5<y<0.5$ (mid-rapidity), and $0.5<y<1$ (forward rapidity).
Each track was required to have hits on more than 15 (out of 45 maximum) of
the rows of TPC readout pads to ensure good tracks.  
A requirement on the distance of closest approach (DCA) to the primary
vertex, $DCA<3$ cm, was imposed to reduce contributions from non-primary 
pions.

  Particle identification is accomplished by measuring energy loss in the
gas, $dE/dx$, for each track and comparing to the expected value for each species 
($i = e^{\pm}$,$\pi^{\pm}$,$k^{\pm}$,$p$,$\bar{p}$) using the equation

\begin{figure}
\includegraphics[width=3.4in]{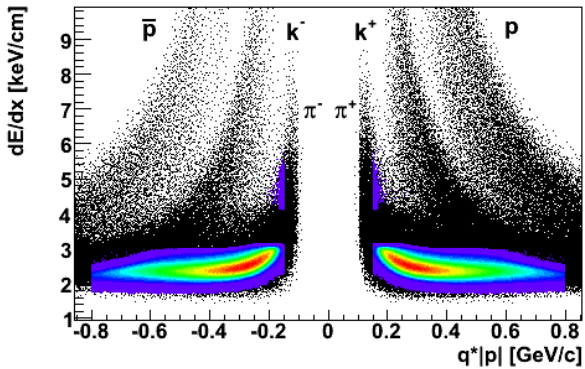}
\caption{\label{F1} (Color online) The energy loss in the TPC, $dE/dx$. The colored
region highlights the pions selected for this analysis.  The gaps in the colored region at 
$|p|\approx0.2$ GeV/$c$ are caused by the cut to eliminate electrons from the 
analysis in the region where the electron and pion bands overlap.  This example
is from $0$-$5\%$ central, 27 GeV Au+Au collisions.}
\end{figure}

\begin{equation}
n\sigma_{i}=\frac{1}{\sigma_{i}}\log \left(\frac{dE/dx_{\mathrm{measured}}}{dE/dx_{\mathrm{expected},{i}}}\right)
\end{equation}
where $\sigma_{i}$ is the $dE/dx$ resolution of the TPC.
Tracks with $|n\sigma_{\pi}|<2$ allow identification of pions for use in the 
analysis.  An additional requirement that $|n\sigma_{e}|$, $|n\sigma_{k}|$, and 
$|n\sigma_{p}| >2$ supresses contamination from other particles.  Additionally, 
a transverse momentum cut, $0.15 < p_{T} < 0.8$ GeV/$c$, further ensures particles 
come from the region where the pion band is separated from the kaon band.  Any
contamination is estimated to be less than $1.7\%$ even before the $n\sigma$ 
cut to reject kaons.  Figure~\ref{F1} demonstrates that these cuts effectively 
remove particles other than pions.  

\subsection{Pair $k_{T}$ cuts and binning}\label{S3d}

  Similar to previous analyses \cite{PRCstarAuAu200,PRLstarAziAuAu200,
CuCuAuAu,ppFemto} pairs were required to have average 
transverse pair momenta, $k_{T}=|\vec{p}_{\mathrm{T1}}+\vec{p}_{\mathrm{T2}}|/2$, in the 
range $0.15 < k_{T} < 0.6$ GeV/$c$.  For the non-azimuthal HBT analyses four
$k_{T}$ bins were used: [0.15,0.25] GeV/$c$, [0.25,0.35] GeV/$c$, [0.35,0.45] GeV/$c$,
[0.45,0.6] GeV/$c$.  This binning allows the presentation of results as a function 
of mean $k_{T}$ (or $m_{T}=\sqrt{k^{2}_{T}+m^{2}_{\pi}}$) in each bin.  These
bins yield mean $k_{T}$ values similar to those in the data from previous
analyses 
allowing direct comparison of certain quantities to previously observed trends.

  In earlier azimuthal HBT studies by CERES \cite{CERESaziHbtPRC} 
and STAR \cite{PRLstarAziAuAu200} the analysis was 
performed in similar, narrow $k_{T}$ bins.  For an azimuthally differential 
HBT analysis the statistics are spread across at least four additional azimuthal
bins.  At the lowest energies this did not allow for sufficient statistics.  
For instance, the 7.7 GeV dataset has both the fewest number of events
and the lowest multiplicity per event in each centrality bin.  Reliable results
could not be obtained from data split into both multiple $k_{T}$ and multiple
bins relative to the reaction plane.  Instead, a single $k_{T}$-integrated 
analysis was performed using all pairs in the combined range 
$0.15 < k_{T} < 0.6$ GeV/$c$ with $\langle k_{T}\rangle\approx$ 0.31 GeV/$c$.  The eccentricity at kinetic freeze-out exhibits 
a systematic decrease by as much as 0.02 when using a single wide $k_{T}$ range compared
to analyses where results from several narrow $k_{T}$ ranges are averaged.
This is simply because the lowest $k_{T}$ bin appears to give a slightly smaller
eccentricity.  
When a wide bin is used the results are biased toward 
the low $k_{T}$ results due to the much higher statistics of the low $k_{T}$ 
pairs.  
In the earlier analyses, CERES reported a weighted average of results for 
different $k_{T}$ bins, while STAR used an average without statistical weights.
In any case, to compare the present results as a function of $\sqrt{s_{NN}}$
the same $k_{T}$ integrated range was used for all energies.

  For the azimuthally differential analysis, the pairs were separated into
four $45^{\circ}$ wide azimuthal bins relative to the reaction plane direction
using the angle $\Phi=\phi_{\mathrm{pair}}-\psi_{2}$. The angle of each pair, 
$\phi_{\mathrm{pair}}$, is the azimuthal angle of the average pair transverse momentum 
vector, $\vec{k}_{T}$, and $\psi_{2}$ is the second-order event plane angle
defined in the range $[0,\pi]$.  This allows measurement of the oscillations 
of parameters necessary to estimate the freeze-out eccentricity as projected
on the transverse plane.  A first order analysis could provide additional
information at the lowest energies \cite{ShapeAnalysis,CorrespondenceHBT}.  
However, significant additional work is needed to obtain first order results 
due to complications from relatively low statistics spread across more bins and 
with much lower first order (compared to second order) event plane resolutions.

\section{Analysis Method}\label{S4}


\subsection{The correlation function}\label{S4a}

  The experimental correlation function is constructed by forming the
distributions of relative momenta, $\vec{q}=(\vec{p}_{1}-\vec{p}_{2})$.  A 
numerator, $N(\vec{q})$, uses particles from the same event, while a mixed event 
denominator, $D(\vec{q})$, uses particles from different events.  The numerator
distribution is driven by two-particle phase space, quantum statistics, and
Coulomb interactions, while the denominator reflects only phase space effects.
Since quantum statistics and final state interactions are driven by freeze-out
geometry \cite{AnnRevHbt}, the ratio 
\begin{equation}
\label{E1b}
C\left(\vec{q}\right)= \frac{N\left(\vec{q}\right)}{D\left(\vec{q}\right)}
\end{equation}
carries geometrical information. 
In the azimuthally differential
analysis, four correlation functions were formed corresponding to four
$45^{\circ}$ wide angular bins relative to the event plane centered at
$0^{\circ}$ (in-plane), $45^{\circ}$, $90^{\circ}$ (out-of-plane), and 
$135^{\circ}$.  The angle between the transverse momentum for each pair and 
the event plane is used to assign each pair to one of the correlation functions.  
The denominators were constructed with pairs formed from mixed events.  Events
were mixed only with other events in the same centrality bin and with relative z
vertex positions of less than 5 cm.
For the azimuthally differential case, events were also required to have the
estimated reaction plane within $22.5^{\circ}$, similar to an earlier analysis
\cite{PRLstarAziAuAu200}.  Reducing the width of the mixing bins only changes
the relative normalizations in the different angular bins but has no effect on
the other fit parameters.  The correlation functions in this analysis are formed
with like-sign pions and the separate distributions for $\pi^{+}\pi^{+}$ and
$\pi^{-}\pi^{-}$ are later combined before fitting since no significant 
difference between the two cases has been observed.

Detector inefficiency and acceptance effects 
apply to both the numerator and denominator and so, in taking the ratio to form the
correlation function, these effects largely cancel.
However, two particle reconstruction inefficiencies allow track splitting and
merging effects which are removed as will be described. 



A single charged particle track may be reconstructed as two tracks with nearly
identical momentum by the tracking algorithm.  This so called track splitting
can strongly affect correlation measurements by contributing false pairs to 
the correlation function at small relative momenta, the signal region.  
The same algorithm, described in \cite{PRCstarAuAu200}, to remove split tracks
is used in the current analysis.  Studies analogous to
those in \cite{PRCstarAuAu200} show the same ``splitting level'' requirement,
$SL<0.6$, is also effective at removing track splitting effects in the current 
data sets.  




On the other hand, two particles with small relative momenta can be reconstructed
as a single track, thus reducing the measured number of correlated particles.
In the following we briefly recapitulate the technique applied for removing
track merging effects, detailed in Ref. \cite{PRCstarAuAu200}.
If two tracks have hits on the same row of readout pads in the TPC that are too close
together, they would appear as a single ``merged'' hit.  Two tracks with 
such ``merged'' hits on many of the 45 rows of TPC readout pads are more likely to be 
reconstructed as a single merged track.  For each pair of tracks, the fraction 
of hits that are close enough so they would appear merged is computed.
The allowed fraction of merged hits (FMH) can be reduced until 
the effect is eliminated.  
The same algorithm can be applied to track pairs from the numerator and
denominator.  
It was determined that 
$\mathrm{FMH}<10\%$ reduced track merging effects as much as possible.  While this 
approach eliminates the potentially large effect of track merging, it introduces 
a systematic uncertainty due to the non-Gaussianess of the correlation function. 
The azimuthal HBT analysis is more sensitive to the 
track merging cut and allows the systematic uncertainty associated
with this requirement to be estimated in Sec.\ref{S4i}.
Analogous studies to those in \cite{PRCstarAuAu200} using current low energy data sets 
lead to the same dependence of the radii on FMH, so in 
the present analyses the same requirement that $\mathrm{FMH}<10\%$ is imposed to remove 
effects of track merging.

\subsection{Bertsch-Pratt parameterization}\label{S4b}

  The relative pair momentum, $\vec{q}$, is projected onto the Bertsch-Pratt 
\cite{BPcoordinatesPratt,BPcoordinatesBertsch,BPcoordinatesRolcrossterm}, 
out-side-long (or o-s-l), coordinate system so that $q_{\mathrm{out}}$ lies along the 
direction of the average transverse pair momentum, $\vec{k}_{T}$, while 
$q_{\mathrm{long}}$ lies along the ``longitudinal'' beam direction, and $q_{\mathrm{side}}$ is 
perpendicular to the other directions and is therefore also in the transverse 
plane.  The relative momentum is expressed in the longitudinal co-moving system 
(LCMS) in which the longitudinal component of the pair velocity vanishes.

  To extract the bulk shape of the particle emitting regions, a Gaussian
parameterization is typically used:

\begin{equation}
\label{E2}
\begin{split}
C & \left(\vec{q}\right)= \left(1-\lambda\right) + K_{\mathrm{Coul}}(q_{\mathrm{inv}})\lambda \\
& \times\exp\left(-q^{2}_{o}R^2_{o}-q^{2}_{s}R^{2}_{s}-q^{2}_{l}R^{2}_{l}-2q_{o}q_{s}R^{2}_{os}-2q_{o}q_{l}R^{2}_{ol}\right)
\end{split}
\end{equation}

\begin{figure}[ht]
\includegraphics[width=3.4in]{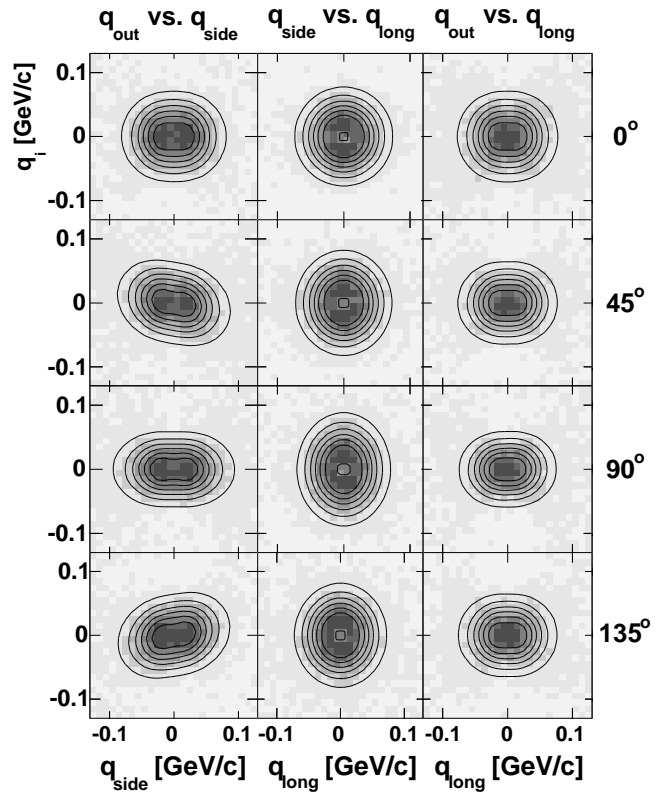} 
\caption{\label{F2} Two dimensional projections of a correlation
function in the $q_{o}$-$q_{s}$, $q_{s}$-$q_{l}$ and $q_{o}$-$q_{l}$ planes for like-sign
pions at mid-rapidity in $20$-$30\%$ central, 27 GeV collisions with
$0.15<k_{T}<0.6$ GeV/$c$.  All scales are in
GeV/$c$.  In each case the third component is projected over $\pm$ 0.03 GeV/$c$.  The
emission angles relative to the event plane are within $\pm22.5^{\circ}$ of the bin
centers indicated along the right side.  The tilt in the $q_{o}$-$q_{s}$ plane is
clearly visible.  Contour lines represent projections of the corresponding fit.}
\end{figure}

The $\lambda$ parameter accounts for non-primary particles that may come
from resonance decays and misidentified particles \cite{PRCstarAuAu200}.
The values of $K_{\mathrm{Coul}}$ account for the Coulomb interaction as discussed in 
the next section.  An overall normalization of the correlation function, 
also determined during the fitting procedure, scales the correlation 
function to a value of unity at large values of $|\vec{q}|$.

  The $R^{2}_{ol}$ term vanishes at mid-rapidity, but becomes positive 
(negative) at forward (backward) rapidity in both azimuthal and non-azimuthal 
analyses \cite{BPcoordinatesRolcrossterm}.  
For the azimuthally integrated analysis $R^{2}_{os}$ vanishes, while in an 
azimuthally differential analysis a second order sinusoidal variation appears 
relative to the reaction plane.  Parametrically, a non-zero cross term 
corresponds to a tilt of the correlation function in $\vec{q}$-space.  This can be 
seen clearly in Fig.~\ref{F2} in the $q_{\mathrm{out}}$-$q_{\mathrm{side}}$ plane.  
At $45^{\circ}$ there is a tilt resulting in a positive $R^{2}_{os}$ cross term.  
At $135^{\circ}$ there is an opposite tilt corresponding to a negative 
$R^{2}_{os}$ cross term.  The interplay between the cross terms and the 
inherent non-Gaussianess of the correlation function is discussed later in an
appendix, where folding the relative momentum distributions allows covariations 
in the fit parameters that would strongly affect the results.  In this analysis,
no folding of $\vec{q}$-space is performed, eliminating this effect.

  In the azimuthally differential analysis, several correlation functions are
constructed for different angular bins.  These are each fit with Eq.~\ref{E2} 
to extract the fit parameters.  The relationship between these fit parameters
describing the regions of homogeneity and the shape of the source region
(the collision fireball at kinetic freeze-out) has been described in several references, such as
\cite{LisaRetiere,CorrespondenceHBT,CorrectionAlgoAndSymmetries}, for boost 
invariant systems.

\subsection{Coulomb interaction}\label{S4c}

  Particles that are nearby in phase space and carry the signal in the
correlation function will also experience Coulomb interactions.  This effect
must be taken into account when extracting the HBT radii.
Different methods of accounting for the Coulomb interaction were studied
systematically in \cite{PRCstarAuAu200}.  
This analysis uses the Bowler-Sinyukov method
\cite{BScoulombBowler,BScoulombSinyukov}.  
The Coulomb interaction is computed for each pair with relative 
momentum components, $(q_o, q_s, q_l)$, that enters the analysis.  The average
interaction in each $(q_o, q_s, q_l)$ bin is included as a constant, $K_{\mathrm{Coul}}$, 
in the fit parameterization.  The quantity $K_{\mathrm{Coul}}$ is the squared Coulomb wave function 
integrated over the entire spherical Gaussian source.  The same radius, 5 fm, 
is used as in earlier analyses.  In Eq.~\ref{E2}, $K_{\mathrm{Coul}}$ only applies to 
the pairs nearby in phase space (the exponential term) and not to other 
particles accounted for by the $(1-\lambda)$ term.  

In principle, correction for the Coulomb interaction between each particle and 
the mean field could also be taken into account.  However, at the energies 
studied here, this interaction has
been found to be negligible \cite{CoulombPiNucleusRef1,CoulombPiNucleusRef2}.

\subsection{Event plane calculations}\label{S4d}

   The azimuthal analysis requires determining the event plane for each event, 
including applying appropriate methods to make the event plane distribution
uniform 
\cite{FlowAnalysisResolution}.  Uncertainty in the event plane reduces the 
extracted oscillation amplitudes of the HBT radii.  The event plane 
resolutions must be computed in order to correct for this effect later in the 
analysis.  The $n^{th}$ order event plane angle, $\psi_{n}$, is determined using 
charged particles measured in the TPC according to the equation

\begin{figure}
\includegraphics[width=3.4in]{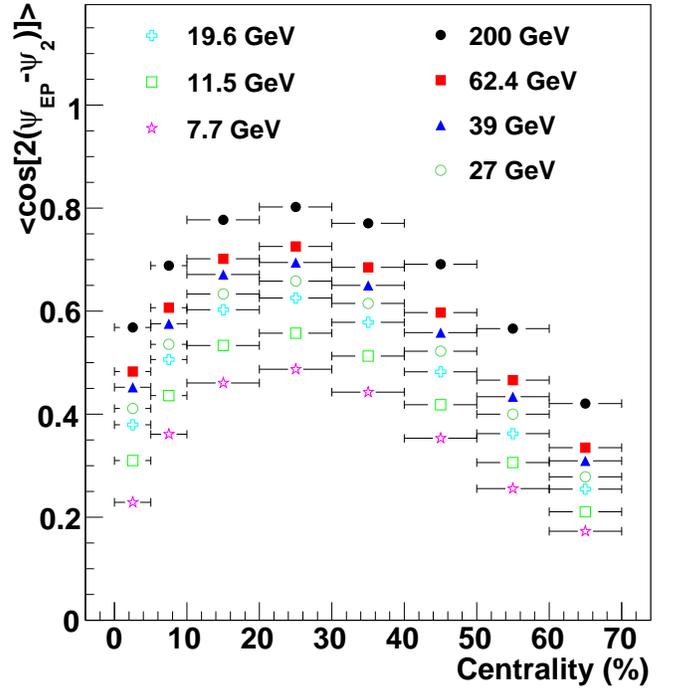} 
\caption{\label{F3}  (Color online) The event plane resolutions for Au+Au collisions at $\sqrt{s_{NN}}$ = 7.7, 11.5, 19.6, 27, 39, 62.4 and 200 GeV as a function of collision centrality.  The resolutions, computed using the TPC ($|\eta|<1$), enter into both the correction algorithm and the global fit method.  Statistical errors are smaller than the symbols.}
\end{figure}

\begin{equation}
\psi_{n} = \frac{1}{n}\arctan\left(\frac{Q_{y}}{Q_{x}}\right)+\Delta\psi_{n}
\end{equation}
where the components of the event plane vector are

\begin{equation}
Q_{x} = \frac{1}{N}\displaystyle\sum_{i} \left(w_{i}\cos(n\phi_{i}) - \langle Q\rangle_{x}\right)
\end{equation}

\begin{equation}
Q_{y} = \frac{1}{N}\displaystyle\sum_{i} \left(w_{i}\sin(n\phi_{i}) - \langle Q\rangle_{y}\right).
\end{equation}
Here, $\phi_{i}$ is the angle of the $i^{th}$ track and $N$ is the total
number of tracks used to determine the event plane.  The shift correction 
\cite{FlowAnalysisResolution} is given by

\begin{equation}
\begin{split}
\Delta\psi_{n}& = \displaystyle\sum_{\alpha=1}^{\alpha_{\mathrm{max}}} \frac{2}{\alpha}(-\langle\sin(n\alpha\psi_{n})\rangle\cos(n\alpha\psi_{n}) \\ 
& \qquad \qquad \; +\langle\cos(n\alpha\psi_{n})\rangle\sin(n\alpha\psi_{n}))
\end{split}
\end{equation}
where $\alpha$ determines the order ($n\alpha$) that each correction term 
flattens.  This analysis is performed relative to the second-order ($n=2$) 
event plane.

  For 7.7-39 GeV, the $\phi$-weighting method
\cite{FlowAnalysisResolution,EPflatteningPWA,EPflatteningPWandShiftA,EPflatteningPWB} was used 
to flatten the event plane.  The inverse, single particle, azimuthal 
distribution is used to weight each particle in the event plane determination 
so that inefficiencies do not affect the event plane determination.  
The $\phi$-weight, $\phi_{\mathrm{wgt,i}}$, is selected from this distribution for 
the $i^{th}$ particle using the direction of the particle's transverse 
momentum vector, $\vec{p}_{T,i}$.
In this case $w_{i} = \phi_{\mathrm{wgt,i}} \cdot p_{T,i}$ while the 
recentering terms $\langle Q \rangle_{x}$ and $\langle Q \rangle_{y}$, as well 
as the shift term $\Delta\psi_{n}$, are all zero.  

For 62.4 and 200 GeV a problematic sector of the TPC was turned off causing a 
rather non-uniform azimuthal distribution.  In this case the recentering and 
shift methods 
\cite{FlowAnalysisResolution,EPflatteningShiftA,EPflatteningPWandShiftA} were 
required to determine the 
event plane accurately.  In this case, $\phi$-weights were not applied so 
$w_{i} = p_{T,i}$.  Here, the average offset in the direction of the $p_{T}$ 
weighted flow vector, $\vec{Q}$, is used to compute $\langle Q\rangle_{x}$ and 
$\langle Q\rangle_{y}$.  After this correction is applied, a shift method is 
needed to correct the event plane values for effects due to other harmonics.  
The shift term $\Delta\psi_{n}$ is determined by computing the correction terms 
$\langle\sin(n\alpha\psi_{n})\rangle$ and $\langle\cos(n\alpha\psi_{n})\rangle$ 
from $\alpha=1$ up to $\alpha=20$ terms, although generally $\alpha_{\mathrm{max}}=2$ 
would be sufficient for a second order analysis \cite{FlowAnalysisResolution}.


  The event plane resolution, 
$\langle\cos[2(\psi_{\mathrm{EP}}-\psi_{2})]\rangle$, due to differences between
the reconstructed ($\psi_{\mathrm{EP}}$) and actual ($\psi_{2}$) reaction planes, 
is also needed as it enters the correction algorithm described later.  The 
calculation begins by determining two event planes for two independent 
subevents which in this analysis correspond to the $\eta<0$ and $\eta>0$ 
regions, so called $\eta$ subevents.  These subevent plane estimates are 
processed through an iterative procedure to solve for the full event plane 
resolution as outlined in \cite{FlowAnalysisResolution}.  Resolutions are 
reduced for lower multiplicity (and therefore lower energy) as well as more 
round (less anisotropic) cases.  The values at each energy that enter this 
specific analysis are included in Fig.~\ref{F3}.

\subsection{Systematic uncertainties}\label{S4i}

  The sources of systematic uncertainty have been studied in previous
HBT analyses such as \cite{PRCstarAuAu200,PRLstarAziAuAu200,CuCuAuAu,ppFemto}.  
Similar studies have been used to estimate the systematic uncertainty due to 
the Coulomb correction, fit range, and fraction of merged hits (FMH) cut discussed 
earlier.  The azimuthal analysis is most sensitive to the fraction of merged hits
requirement and this is used to estimate the systematic uncertainty.  For lower energies
the dependence of the fit parameters on the allowed fraction of merged hits is
consistent with earlier results at $\sqrt{s_{NN}} = 200$ GeV. 
Reduction of the Coulomb radius from 5 fm to 3 fm
and variation of the fit range from 0.15 GeV/$c$ to 0.18 GeV/$c$, also leads to results similar to earlier studies.
Track splitting is effectively eliminated.  
The uncertainties are estimated to be the same for each $\sqrt{s_{NN}}$ reported
here and are summarized in Table II, for each source, for the HBT radii and
freeze-out eccentricity (defined in Sec.~\ref{S5b3}).

\begin{table}
\begin{center}
\begin{tabular}{lllll}
\hline \hline
\\[3pt]
Source            &       $R_{\mathrm{out}}$        &       $R_{\mathrm{side}}$    &    $R_{\mathrm{long}}$    &   $\varepsilon_{F}$\\[3pt]
\hline
\\[3pt]
Coulomb         &       4\%       &       3\%       &       4\%      &      0.004 \\
Fit Range       &       5\%       &       5\%       &       5\%      &      0.002 \\
FMH             &       7\%       &       3\%       &       3\%      &      0.003 \\
\hline
Total           &     9.5\%       &     6.5\%       &       7\%      &      0.005 \\[3pt]
\hline \hline
\end{tabular}
\caption{\label{T2} The approximate systematic uncertainty on the HBT radii and freeze-out eccentricities.}
\end{center}
\end{table}
Earlier STAR analyses \cite{PRCstarAuAu200,PRLstarAziAuAu200,CuCuAuAu,ppFemto}
found, for various collision species (p+p, Cu+Cu, Au+Au) and
data sets that the systematic uncertainty is approximately $10\%$ or less for the HBT
radii in all centrality and $k_{T}$ bins studied.  Analogous studies lead
to the same conclusion for the data sets used in the current analysis and suggest
the uncertainties are virtually independent of beam energy.

  It should be noted that there is also an inherent uncertainty in the general 
method used to extract the eccentricity.  
The theoretical framework assumes a static, Gaussian region of homogeneity 
that corresponds to the entire volume of the collision at kinetic freeze-out.  
Flow-induced space-momentum correlations reduce this correspondence 
which could affect the reliability of the equations.
However, several different model studies
\cite{LisaRetiere,CorrespondenceHBT} find 
consistently that the results are still reliable to within $30\%$, even in the
presence of strong flow.  This would not affect any conclusions regarding the
shape of the excitation function in regards to whether or not it is monotonic.

\section{Extracting radius oscillations}\label{Sb4}

  In azimuthally differential analyses, correlation functions are constructed
for pairs directed at different angles relative to the event plane.  The HBT
radii that describe these regions exhibit sinusoidal variations relative to the
event plane direction.  Second order oscillations of these radii can be described
in terms of Fourier coefficients which have been related to the eccentricity of
the collision fireball at kinetic freeze-out.  Due to finite-bin-width and event
plane resolution, the amplitude of these oscillations is reduced from the
actual value.  In order to determine the true amplitudes, these effects must be
taken into account.  
Three methods of correcting for these effects will be described later in this
section.


\begin{figure}
\includegraphics[width=3.4in]{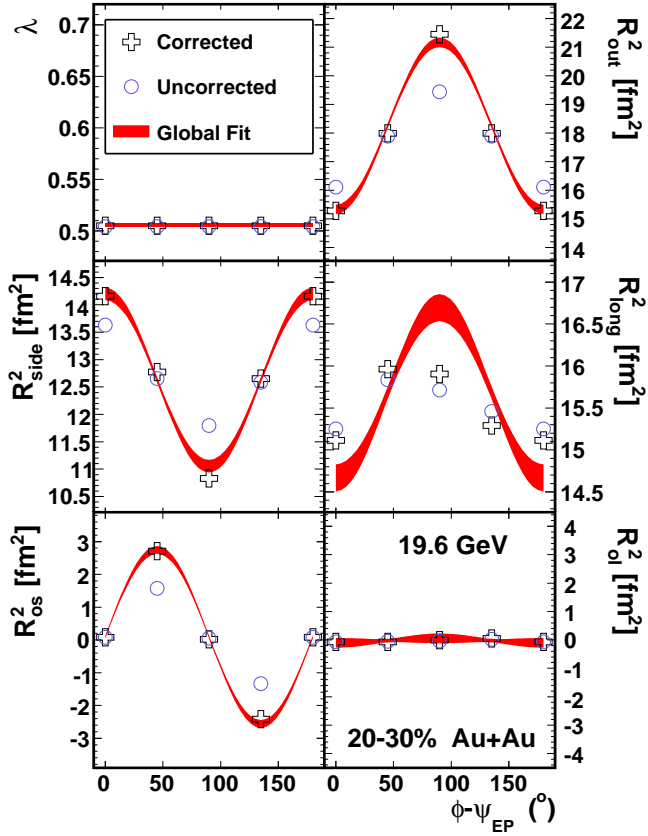} 
\caption{\label{F4}  (Color online) Examples of the angular oscillations of the HBT radii relative to the event plane from $20$-$30\%$ central, 19.6 GeV Au+Au collisions for $0.15<k_{T}<0.6$ GeV/$c$.  Open circles show the radii before correction for finite-bin-width and event plane resolution.  Open cross symbols demonstrate that correcting these effects increases the oscillation amplitude.  The corrected and uncorrected results are obtained with the HHLW fit method (see text) before and after the correction algorithm (Sec.\ref{Sb4f}) is applied.  The points at $0^{\circ}$ are repeated on the plot at $180^{\circ}$ for clarity.  The solid band shows the Fourier decomposition directly extracted using a global fit (Sec.~\ref{Sb4c}) to all four angular bins.  The value of $\lambda$ is consistent for the two methods.}
\end{figure}

  In the azimuthal HBT analysis, four correlation functions are constructed,
for pairs directed in four different angular bins centered at $\Phi$ =
$0^{\circ}$, $45^{\circ}$, $90^{\circ}$, and $135^{\circ}$ relative to event plane.  
This allows extraction of the second order sinusoidal variations of the HBT 
radii.
Figure~\ref{F4} shows an example of these oscillations.  
The $\Phi$ dependence of the HBT radii for a given beam energy, centrality,
and $k_{T}$ are described by:

\begin{equation}\label{E7}
\begin{split}
R^{2}_{\mu}&\left(\Phi\right) = R^{2}_{\mu,0} \\
& +2\displaystyle\sum_{n=2,4,6...} R^{2}_{\mu,n} \cos(n\Phi) \qquad \left(\mu=o,s,l,ol\right)
\end{split}
\end{equation}
and
\begin{equation}\label{E8}
\begin{split}
R^{2}_{\mu}&\left(\Phi\right) = R^{2}_{\mu,0} \\
& +2\displaystyle\sum_{n=2,4,6...} R^{2}_{\mu,n} \sin(n\Phi) \qquad \left(\mu=os\right) \quad \; \: 
\end{split}
\end{equation}
where $R^{2}_{\mu,n}$ are the $n^{\mathrm{th}}$-order Fourier coefficients for
radius term $\mu$.  These coefficients are computed using

\begin{equation}\label{E9}
R^{2}_{\mu,n} = \begin{cases}
\langle R^{2}_{\mu}\left(\Phi\right)\cos(n\Phi)\rangle \qquad \left(\mu=o,s,l,ol\right) \\
\langle R^{2}_{\mu}\left(\Phi\right)\sin(n\Phi)\rangle \qquad \left(\mu=os\right)
\end{cases}
\end{equation}
The $0^{\mathrm{th}}$-order Fourier coefficients are expected to be nearly identical
to radii extracted in an azimuthally integrated analysis.  The $2^{\mathrm{nd}}$-order 
terms correspond to half the amplitude of the second order oscillations for a 
second order, $n=2$, analysis.  


  Imperfect event plane resolution smears the difference between neighbouring 
azimuthal bins and it also causes the peaks of the extracted oscillations 
of the HBT radii to appear smaller 
than they ideally should be.  These effects must be corrected for in order to 
extract the true $2^{\mathrm{nd}}$-order oscillation amplitudes needed to
compute the kinetic freeze-out eccentricities, $\varepsilon_{F}$, which are
discussed in Sec.~\ref{S5b3}.  In the following discussion, two methods that have 
been applied in earlier analyses (which we refer to as the E895 and HHLW 
methods) are reviewed.  A third method used in this analysis, dubbed the global 
fit method, is then introduced.

%

\subsection{E895 method}\label{Sb4a}



  In an earlier azimuthal HBT analysis performed by the E895 collaboration 
\cite{E895aziHbtData} and a later analysis by the CERES collaboration
\cite{CERESaziHbtPRC}, the radii were extracted from correlation functions that 
were uncorrected for finite-bin-width and resolution effects.  These uncorrected 
radii were then used to compute the Fourier coefficients described above.  The 
uncorrected, $2^{nd}$-order Fourier coefficients were then scaled by dividing by 
the event plane resolution, as is done when correcting a $v_{2}$ measurement for
event plane resolution effects.  While this is found to give consistent results to other
methods described below, it is formally incorrect because it is not the radii 
that are smeared, but rather each $q$-space bin for each of the numerator,
denominator and Coulomb weighted mixed event distributions separately.  This
method will be referred to as the E895 method.

\subsection{HHLW method}\label{Sb4f}

  In this method, used first in \cite{PRLstarAziAuAu200}, a model independent
correction algorithm is applied to compute the corrected numerator,
denominator, and Coulomb weighted denominator histograms for each angular bin. 
The radii extracted from these corrected distributions are then used to 
compute the Fourier coefficients.  
This method will be referred to as the HHLW method after the
authors of the paper in which it was developed
\cite{CorrectionAlgoAndSymmetries}.  We briefly summarize this correction 
procedure below.

  The derivation, detailed in Ref.~\cite{CorrectionAlgoAndSymmetries}, requires first 
decomposing mathematical expressions for the true (corrected) and experimental 
distributions as Fourier series.  The
true distributions are then convolved with a (Gaussian) distribution of the 
reconstructed event plane centered about the reaction plane, and a function to account
for the finite azimuthal bin width.  Finally, each coefficient from the series
for the true distribution is equated with the corresponding coefficient from the
series expansion of the experimental distribution.  
This leads to the following relationship between coefficients for the
true and experimentally observed distributions:

\begin{equation}\label{E13a}
A^{\mathrm{exp}}_{\alpha,n}\left(\vec{q}\right)=A_{\alpha,n}\left(\vec{q}\right)\frac{\sin(n\Delta/2)}{n\Delta/2}\langle\cos[n(\psi_{\mathrm{EP}}-\psi_{2})]\rangle.
\end{equation}
The quantities $A_{\alpha,n}(\vec{q})$ and $A^{\mathrm{exp}}_{\alpha,n}(\vec{q})$
are the coefficients for the Fourier series representation of the true and
experimental distributions respectively.  
The formula applies separately to the numerator ($A$=$N$) and the denominator 
($A$=$D$) of Eq. \ref{E1b}
and the Coulomb weighted mixed event ($A$=$K_{\mathrm{Coul}}$) distributions.
The factors
multiplying $A_{\alpha,n}(\vec{q})$ come from the convolution of the true series
mentioned previously.  The quantities $\langle\cos[n(\psi_{\mathrm{EP}}-\psi_{2})]\rangle$
are the reaction plane resolutions. 
The symbol $\Delta$ is the
width of each angular bin and $n$ is the order of the Fourier coefficient.  The
experimental coefficients can be computed from the experimentally measured
distributions in each angular bin using the standard definition for Fourier
coefficients so that

\begin{equation}\label{E13b}
A^{\mathrm{exp}}_{\alpha,n}(\vec{q}) = \begin{cases}
\langle A^{\mathrm{exp}}_{n}(\vec{q},\Phi)\cos{n\Phi}\rangle \qquad \left(\alpha=c\right) \\
\langle A^{\mathrm{exp}}_{n}(\vec{q},\Phi)\sin{n\Phi}\rangle \qquad \left(\alpha=s\right)
\end{cases}
\end{equation}
are the coefficients for the cosine ($\alpha=c$) or sine ($\alpha=s$) terms in
the series expansion.
  
  The corrected distributions can be computed from the experimental
distributions using
\begin{equation}
\label{E13}
\begin{split}
A\left(\vec{q},\Phi_{j}\right)=& A^{\mathrm{exp}}\left(\vec{q},\Phi_{j}\right) + 2\sum^{n_{\mathrm{\mathrm{bins}}}}_{n=1}\zeta_{n}\left(\Delta\right) \\
& \times [A^{\mathrm{exp}}_{c,n}\left(\vec{q}\right)\cos(n\Phi_{j}) + A^{\mathrm{exp}}_{s,n}\left(\vec{q}\right)\sin(n\Phi_{j})].
\end{split}
\end{equation}
 In this analysis, only the $2^{\mathrm{nd}}$-order event plane, $\psi_{2}$, is 
measured, and so only the $n=2$ terms are required.  
The correction parameter $\zeta_{n}(\Delta)$ is given by 
\begin{equation}
\label{E14}
\zeta_{n}\left(\Delta\right)=\frac{n\Delta/2}{\sin(n\Delta/2)\langle\cos[n(\psi_{\mathrm{EP}}-\psi_{2})]\rangle}-1.
\end{equation}
Substituting Eq.~\ref{E14} into Eq.~\ref{E13} leads to an identity, with only
experimentally measured quantities on the right hand side.

  Once the corrected numerator, denominator, and Coulomb weighted mixed-event
distributions are computed for each angular bin, fits are performed to
extract the radii.  As in \cite{PRLstarAziAuAu200}, the $\lambda$ parameter from the four angular 
bins are averaged (for each centrality) and set as a constant for all four
bins; the $\langle \lambda \rangle$ values are nearly identical to the non-azimuthal cases.
The correlation functions are refit to extract the radii.  The $\lambda$-fixing
procedure reduces the number of independent fit parameters needed.  This 
procedure is done under the assumption that $\lambda$ has no explicit $\Phi$ 
dependence and to date none has been observed.

  In any case, the HBT radii extracted from these corrected distributions
exhibit the true, larger oscillation amplitude.  This is clearly demonstrated 
in Fig.~\ref{F4}.  One deficiency in this approach is that the uncertainties on the
corrected distributions are correlated, leading to an underestimate of the 
uncertainties for the extracted radii.  We have developed a global fit 
method, 
described next, to avoid this issue.


 
\subsection{Global fit method}\label{Sb4c}

 
  A new global method of fitting was developed that avoids correlated errors
and provides more reliable results in cases of low statistics and poor event 
plane resolution.  
The method 
begins with the same Gaussian parameterization as in Eq.~\ref{E2}.  The Fourier 
representation of the radii from Eqs.~\ref{E7} and~\ref{E8} are substituted, 
keeping only the $0^{\mathrm{th}}$- and $2^{\mathrm{nd}}$-order terms.  In this
method, the fit parameters are the 
Fourier coefficients that describe the oscillations of the radii relative to the
event plane, and so the Fourier coefficients are extracted directly rather than 
the radii.  Using this parameterization, the theoretical estimate of the true 
numerator, $N^{\mathrm{true}}$, is then smeared for event plane resolution 
and finite-binning effects by applying the correction algorithm in reverse, as 
described below.  In this way, a theoretical estimate of the values expected in 
each uncorrected numerator, $N^{\mathrm{smeared}}$, is obtained which 
can then be compared to the uncorrected numerators that are experimentally 
measured, $N^{\mathrm{exp}}$.

\begin{figure*} 
\includegraphics[width=6.0in]{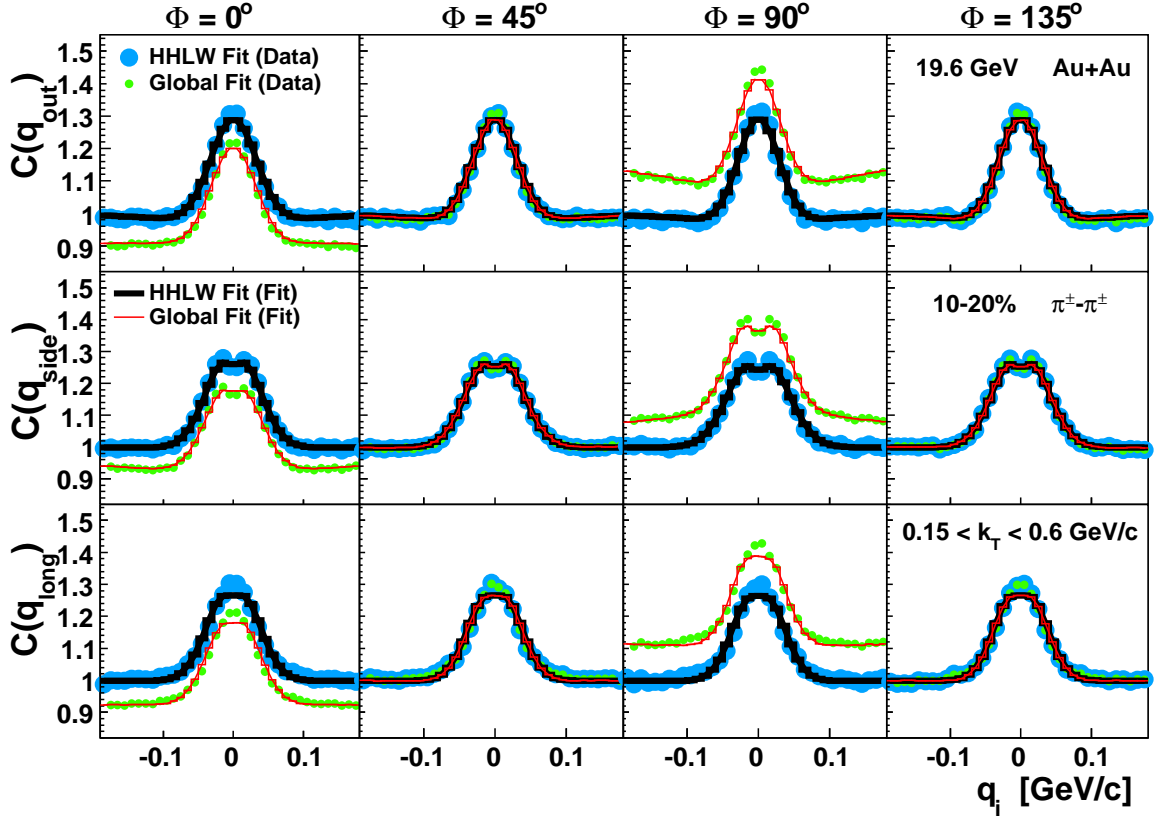} 
\caption{\label{F5}  (Color online) Sample fit projections onto the $q_{\mathrm{out}}$ (top row), $q_{\mathrm{side}}$ (middle row) and $q_{\mathrm{long}}$ (bottom row) axes for four angular bins relative to the reaction plane. Results from the HHLW fit method and the global fit method are shown for direct comparison.  These projections are from results for $10$-$20\%$ central, 19.6 GeV Au+Au collisions with $0.15<k_{T}<0.6$ GeV/$c$.}
\end{figure*}

  For each bin $\vec{q}=$($q_{o}$,$q_{s}$,$q_{l}$), a value of the correlation 
function, $C^{\mathrm{true}}(\vec{q})$, is computed.  An estimate for the denominator 
is obtained from the ``true'' denominator, $D(\vec{q})$ (i.e., the denominator 
for a given $\Phi$ bin run through the correction algorithm described in the 
last section).  The estimate for the true numerator, for each $\vec{q}$ bin, is 
simply $N^{\mathrm{true}}(\vec{q})=D(\vec{q})C^{\mathrm{true}}(\vec{q})$.  This value 
is then run through the correction algorithm in reverse.  A series similar to 
Eq.~\ref{E13}, 

\begin{equation}\label{E15}
\begin{split}
N^{\mathrm{smeared}}&\left(\vec{q},\Phi_{j}\right) = N^{\mathrm{true}}\left(\vec{q},\Phi_{j}\right) + 2\sum^{n_{\mathrm{\mathrm{bins}}}}_{n=1}\zeta^{'}_{n}\left(\Delta\right) \\
& \times[N^{\mathrm{true}}_{c,n}\left(\vec{q}\right)\cos(n\Phi_{j})+N^{\mathrm{true}}_{s,n}\left(\vec{q}\right)\sin(n\Phi_{j})],
\end{split}
\end{equation}
is used to compute the value expected to appear in the uncorrected numerator,
$N^{\mathrm{exp}}$,
for each ($q_{o}$,$q_{s}$,$q_{l}$) bin and each $\Phi$ bin.  The quantity
$N^{\mathrm{smeared}}$ is the value expected in the uncorrected numerator,
$N^{\mathrm{exp}}$, based on the value, $N^{\mathrm{true}}$, predicted by the 
current values of the fit parameters during each iteration of the fit algorithm.  
All fit parameters (including normalizations) obtained in this method 
correspond to the true correlation 
function even though the fit is applied to the uncorrected numerators.  As in 
Eq.~\ref{E13}, only $n=2$ terms are used for an analysis relative to the second 
order event plane.

A factor similar to Eq.~\ref{E14}, from the same relationship between 
true and experimental values, 

\begin{equation}\label{E16}
\zeta^{'}_{n}(\Delta)=\frac{\sin(n\Delta/2)\langle\cos[n(\psi_{\mathrm{EP}}-\psi_{2})]\rangle}{n\Delta/2}-1
\end{equation}
smears the true amplitude according to the resolution and finite-bin-width
when substituted into Eq.~\ref{E15}.

  In this way, an estimate, $N^{\mathrm{smeared}}$, of the value that should be 
found in the uncorrected, raw numerator histogram, $N^{\mathrm{exp}}$, for each 
($q_{o}$,$q_{s}$,$q_{l}$) bin in each
$\Phi$ bin is obtained from the fit function.  The value expected by the fit
function is
compared to the value actually observed in each ($q_{o}$,$q_{s}$,$q_{l}$) bin 
in the four uncorrected numerator histograms for all four $\Phi$ bins in a 
single simultaneous ``global'' fit.


  A separate normalization is used for each $\Phi$ bin since there will be 
differences in the number of tracks, and therefore pairs, in the different
bins.  A single $\lambda$ parameter is used for all four angular bins, as
is done in the HHLW fit method.  The global fit method significantly reduces the 
number of parameters needed 
to describe the data from  
21 parameters ($\lambda$ + 5 radii x 4 $\Phi$ bins) in the HHLW method to 
11 parameters ($\lambda$ + 10 Fourier coefficients), not counting the four normalization parameters.

  The HHLW correction algorithm computes a corrected histogram from
all of the uncorrected histograms.  Therefore, the uncertainties in each 
corrected histogram depend on the uncertainties in all the uncorrected 
histograms.  While the uncertainties are independent in the uncorrected 
histograms, the uncertainties in the ``corrected'' histograms are not.  
However, the fit assumes the uncertainties are independent and, as a result, 
underestimates the true uncertainty.  The new method, by fitting directly to 
the uncorrected numerator histograms, avoids this problem.

  A disadvantage of the new algorithm is that the normalizations obtained
correspond to the ``true'' correlation function, 
$C^{\mathrm{true}}(\vec{q})=N^{\mathrm{true}}(\vec{q})/D^{\mathrm{true}}(\vec{q})$, 
but the fit uses the corrected denominator histogram, $D(\vec{q})$, as in the 
HHLW method, and the uncorrected numerator histogram, 
$N^{\mathrm{exp}}(\vec{q})$.  To compare the fit to the distributions that are 
actually used in the fit, $C'(\vec{q})=N^{\mathrm{exp}}(\vec{q})/D(\vec{q})$ is 
projected onto the out, side and long axes, but the normalizations do not 
correspond exactly.  They do put the 
projections on a common scale however.  The $0^{\circ}$ and $90^{\circ}$ 
projections are shifted away from unity at large $\vec{q}$.  Examples of the 
projections using the global fit method are shown in Fig.~\ref{F5} for the same 
centrality and energy as the fits using the HHLW fit method, also shown in 
Fig.~\ref{F5} for comparison.  As a check, if instead one projects 
$N(\vec{q})/D(\vec{q})$ and $N^{\mathrm{fit}}(\vec{q})/D(\vec{q})$, 
where $N^{\mathrm{fit}}(\vec{q})$ is the unsmeared fit numerator computed from 
the extracted Fourier coefficients (from the global fit method), the projections look essentially identical 
to the HHLW fit method projections for all four angular bins.

  For most centralities and fit parameters, the results agree quite well.
However, the amplitude describing the $R^{2}_{\mathrm{long}}$ oscillation, 
$R^{2}_{l,2}$, is larger when obtained using the new fit method.  This is
demonstrated most clearly in Fig.~\ref{F4} by comparing the solid band for the
oscillation extracted using the global fit method to the corrected radii using
the HHLW method.  The difference 
in $R^{2}_{l,2}$ for the two parameterizations means that the second order
oscillation that best fits the data from all angular bins simultaneously is not
consistent with the Gaussian $R_{\mathrm{long}}$ values that best describe the regions of
homogeneity in each angular bin separately.  The difference may be attributed 
to a subtle interdependence of the fit parameters in the HHLW fit method 
that constrains the $R_{\mathrm{long}}$ values.  Also, the new fit method has 
difficulties in all central $0$-$5\%$ cases and in a few $5$-$10\%$ cases when 
the statistics become low.  These cases are excluded, for instance, from 
Fig.~\ref{F10} as well as all other figures for the azimuthally differential 
analysis.  For some of the $0$-$5\%$ cases the fit could never
converge even with high statistics.  For these unreliable cases, while the 
$R^{2}_{ol,2}$ values are close 
to zero in the HHLW fit results for all centralities, a large $R^{2}_{ol,2}$ 
suddenly appears in this most central bin when using this global fit 
method.  This is likely non-physical 
because, for a symmetric acceptance window around 
mid-rapidity, $R^{2}_{ol}$ must average to zero.  Additionally, because the 
different angular bins are most similar in central events any second order 
oscillation of $R^{2}_{ol,2}$ should decrease in the most central bin due to 
symmetry, not appear suddenly.  In fact when $R^{2}_{ol,2}$ is varied, the 
$\chi^{2}$ value between the fit and the data becomes quite flat for the central 
data compared to other centralities allowing $R^{2}_{ol,2}$ to take on a wide
range of values without constraint.  Additionally, when this happens the 
oscillations extracted for some, or sometimes all, of the other parameters 
($R^{2}_{o,2}$, $R^{2}_{s,2}$, $R^{2}_{l,2}$) change sign in this central case, 
even when statistics are high.  

  Due to the symmetry of the almost round 
central events, the distributions for different angular bins are quite similar 
compared to other centralities.  The global fit method extracts oscillations, 
not radii, from all four bins simultaneously, and when the distributions are 
similar it seems to have the freedom to find a wider variety of solutions.
The HHLW fit method, with separate fits in each azimuthal bin, has no such
freedom, but is found to be less reliable when statistics and resolutions are 
low.  For the global fit method, for other centralities, the results are rather 
stable.  The 
$0^{\mathrm{th}}$-order coefficients remain consistent with the azimuthally
integrated results, which is even true for $0$-$5\%$ centrality.  
The behavior for central data appears to 
be the result of the relationship between the fit parameters used,
the similar shape of the emission regions for all the 
angular bins in the central data, and the very shallow minimum in $\chi^{2}$ that
develops for $R^{2}_{ol,2}$ at the same time.  There are no other differences 
in the global fit algorithm compared to the HHLW fit method.

\section{Results}\label{S5}

  The azimuthally integrated HBT results are discussed first and compared to
historical data from earlier experiments and recent results from ALICE.  Later, 
the azimuthally differential analysis is presented for a wide range of beam 
energies.  The azimuthally differential analysis is also performed in three 
rapidity bins allowing extraction of the excitation function for the 
$R^{2}_{ol}$ parameter and direct comparison of the freeze-out eccentricity in 
the same forward rapidity window as an earlier measurement by the 
CERES collaboration.  Finally, the excitation function for the freeze-out 
eccentricity is discussed along with its implications for the relevant 
underlying physics as outlined in Sec.~\ref{S2}.

\subsection{Azimuthally integrated HBT}\label{S5a}

  There is a wealth of earlier 
HBT data demonstrating the systematic 
behavior of the HBT radii as a function of beam energy, $k_{T}$ (or $m_{T}$), 
and centrality.  Trends have been established despite the measurements 
having been performed by various experiments and with differences in the 
analysis techniques.  In this paper, the results are presented across a wide 
range of beam energies, overlapping previously measured regions and filling 
in previously unmeasured regions of $\sqrt{s_{NN}}$.

\begin{figure}[ht!]
\includegraphics[width=3.4in]{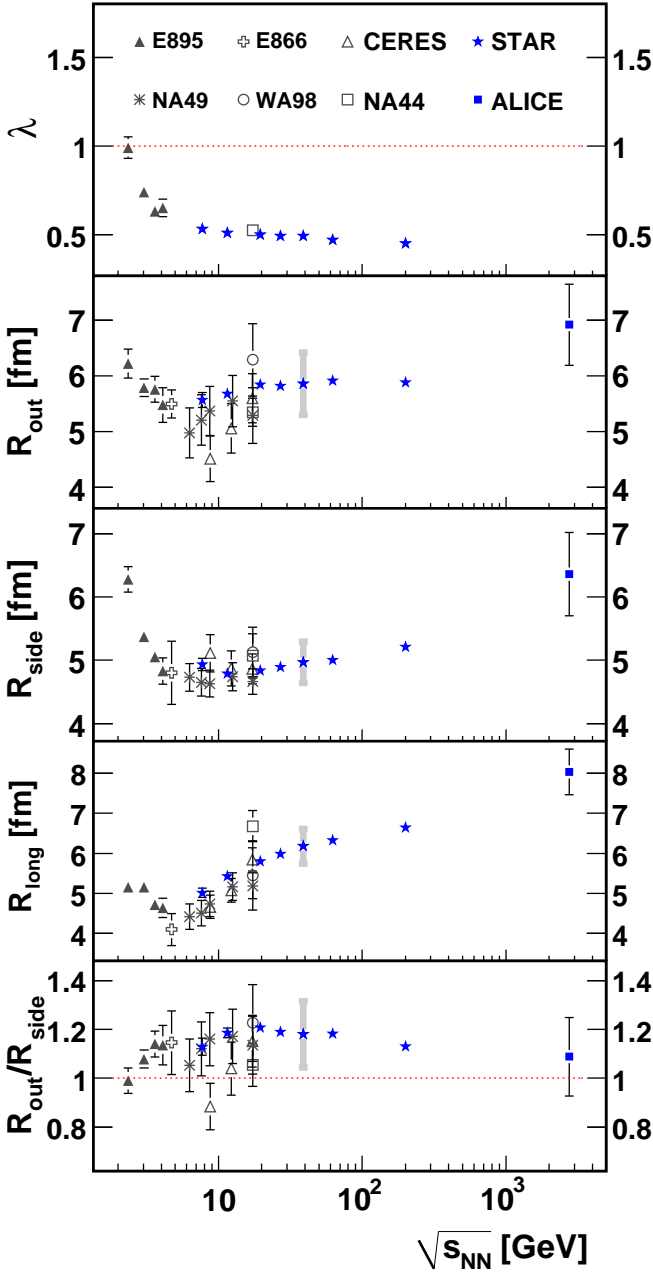} 
\caption{\label{F7}  (Color online) Energy dependence of the HBT parameters for central Au+Au, Pb+Pb, and Pb+Au collisions at mid-rapidity 
and $\langle k_{T}\rangle\approx0.22$ GeV/$c$ \cite{RvsRootSCERES,RvsRootSE895,RvsRootSNA44,RvsRootSNA49,RvsRootSE866,RvsRootSWA98,AliceHbtPLB2010}.  
The text contains discussion about variations in centrality, $k_{T}$, and analysis techniques between experiments.  Errors on NA44, NA49, WA98, CERES, and 
ALICE points include systematic errors.  The systematic errors for STAR points 
at all energies (from Table~\ref{T2}) are of similar size to error bar for 39 GeV, 
shown as a representative example.  Errors on other results are statistical
only,
to emphasize the trend.  For some experiments the $\lambda$ value was not specified.}
\end{figure}

  Figure~\ref{F7} shows the beam energy dependence of the $\lambda$ parameter,
the HBT radii, and the ratio $R_{\mathrm{out}}/R_{\mathrm{side}}$ for like-sign
pions in central
collisions at low $k_{T}$.  All the STAR results are from the most central
$0$-$5\%$ and lowest $\langle k_{T} \rangle$ ($\approx 0.22$ GeV/$c$) data.  
The ALICE point is
also from $0$-$5\%$ central data, but has a slighly larger $\langle k_{T}\rangle \approx 0.26$ 
GeV/$c$.  Results from earlier experiments come from a range of 
central data sets, as narrow as $0$-$7.2\%$ to as wide as $0$-$18\%$ centrality, 
as well as a range of $\langle k_{T}\rangle$ values, from $0.17$ GeV/$c$ to $0.25$ GeV/$c$.
The earlier data are from $\pi^{-}$-$\pi^{-}$ correlation results in which 
various methods of accounting for the Coulomb interaction were employed.  The 
new STAR results are from 
combined $\pi^{-}$-$\pi^{-}$ and $\pi^{+}$-$\pi^{+}$ correlation functions.  
No significant difference between the two cases has been observed so the 
combination simply leads to higher statistics.  Our high-statistics analysis, 
with identical acceptance for all $\sqrt{s_{NN}}$, yields a well-defined smooth 
excitation function consistent with the previous trends.

  The $\lambda$ parameter primarily represents the fraction of correlated
pairs entering the analysis, as described in Sec.~\ref{S4b}.  It decreases with
increasing $\sqrt{s_{NN}}$
relatively rapidly at lower, AGS, energies while changing rather
little from 7.7 to 200 GeV.  This suggests that the fraction of pions in this
$\langle k_{T} \rangle$ range
from long-lived resonances increases at lower energy but remains rather 
constant at higher energies.  The value of $\lambda$ is larger than our earlier 
reported results for 200 GeV~\cite{PRCstarAuAu200} which is related to our
implementation of an anti-electron cut that reduces 
contamination in this analysis.  The $R_{\mathrm{out}}$ parameter similarly shows
little change over a wide range of 
RHIC energies.  It does appear to rise
noticeably at the LHC.  The values of $R_{\mathrm{side}}$ show a very small
increase at the higher RHIC energies and a more significant increase at the
LHC.  The values of $R_{\mathrm{long}}$, on the other hand, appear to reach a minimum around 5
GeV, rising significantly at RHIC and the ALICE point is once again higher 
than the trend observed at STAR.

  The radius $R_{\mathrm{side}}$ is primarily associated with the spatial extent
of the particle emitting region, whereas $R_{\mathrm{out}}$ is also affected by 
dynamics \cite{LisaRetiere,CorrespondenceHBT} and is believed to be related to 
the duration of particle emission \cite{RoRsRatioRef1,RoRsRatioRef2}.  
The ratio $R_{\mathrm{out}}/R_{\mathrm{side}}$ was predicted to increase with 
beam energy by hydrodynamical calculations and might show an enhancement 
if the lifetime of the collision evolution (and, within these models, the 
duration of particle emission as a result) were to be extended 
by entrance into a different phase \cite{RoRsRatioRef1,RoRsRatioRef2}.  
The present measurements reduce statistical fluctuations and fill in the gaps
of the existing excitation function between SPS and top RHIC energies.  The
previous observation that this ratio shows a quite flat energy dependence
is reproduced with the scatter in data points greatly reduced.
The trend remains flat up
to LHC energies.  Model comparisons to this 
trend are discussed in \cite{CuCuAuAu}.

\begin{figure}
\includegraphics[width=3.4in]{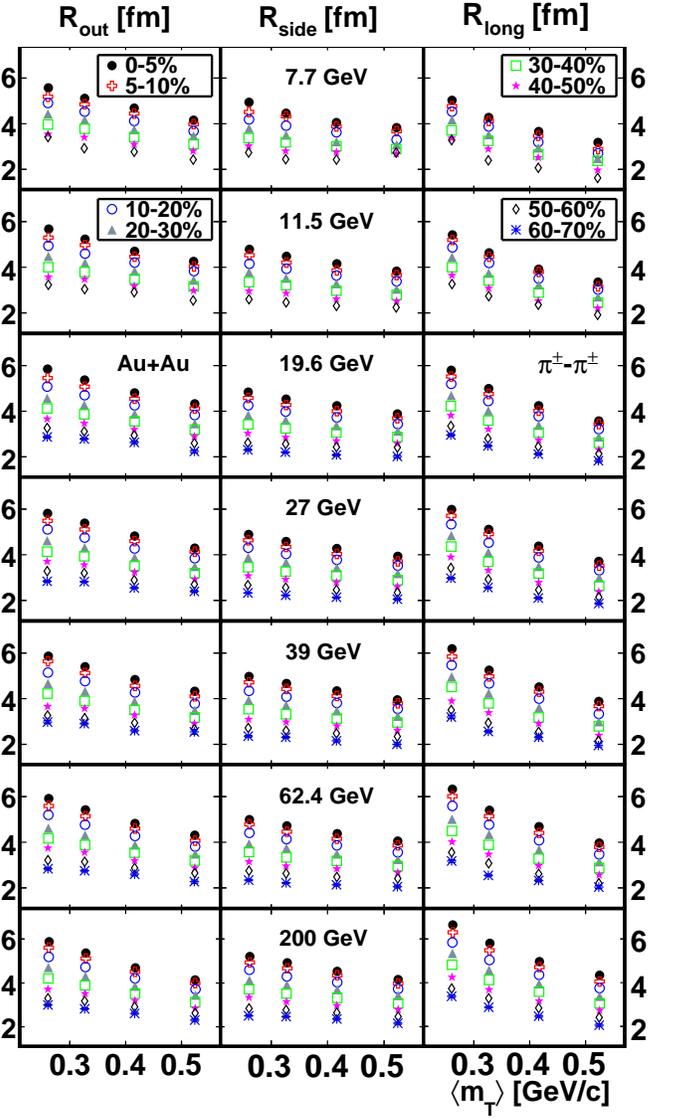} 
\caption{\label{F8}  (Color online) The $\langle m_{T}\rangle$ dependence of $R_{\mathrm{out}}$, $R_{\mathrm{side}}$ and $R_{\mathrm{long}}$ for each energy and multiple centralities.
Errors are statistical only.  For 7.7 GeV and 11.5 GeV, the results for $60$-$70\%$ centrality are excluded due to lack of statistics.}
\end{figure}

\begin{figure}
\includegraphics[width=3.4in]{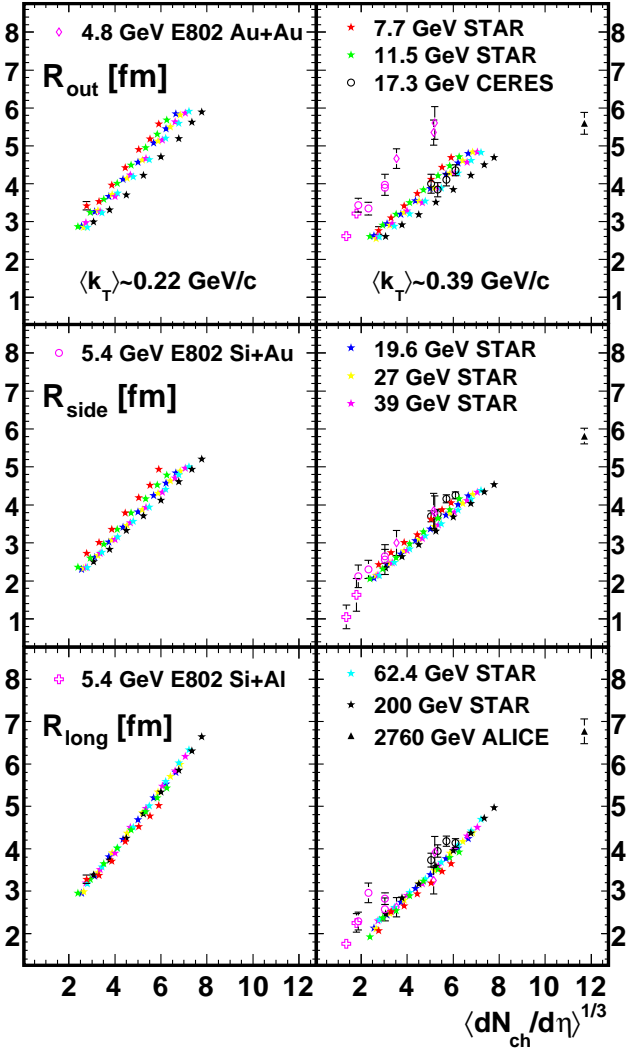} 
\caption{\label{F9}  (Color online) The dependence of the HBT radii on multiplicity, $\langle dN_{\mathrm{ch}}/d\eta\rangle^{1/3}$, for $\langle k_{T}\rangle\approx0.22$ GeV/$c$ (left) and $\langle k_{T}\rangle\approx0.39$ GeV/$c$ (right).  Results are for Au+Au collisions at STAR, Pb+Au at CERES \cite{RvsRootSCERES}, Pb+Pb at ALICE \cite{AliceHbtPLB2010}, and Si+A at E802 \cite{RvsRootSE802}.  Errors are statistical only.}
\end{figure}

  The value of $R_{\mathrm{long}}$ has been related to the kinetic freeze-out 
temperature, $T$, and lifetime, $\tau$, of the system by the relation
\cite{RlongTauRefBertcsh,RlongTauRefHeinz,LisaRetiere}
\begin{equation}\label{E17}
R_{\mathrm{long}} = \tau\sqrt{\frac{T}{m_{T}}\frac{K_{2}(m_{T}/T)}{K_{1}(m_{T}/T)}}
\end{equation}
where $K_{1}(m_{T}/T)$ and $K_{2}(m_{T}/T)$ are modified Bessel functions.
The kinetic freeze-out temperature is not expected to change much with $\sqrt{s_{NN}}$.  
Therefore, the rise of $R_{\mathrm{long}}$ suggests that the total lifetime of the 
system is increasing with energy.  At the end of this section
Eq.~\ref{E17} will be used to extract $\tau$ as a function of $\sqrt{s_{NN}}$
given certain assumptions.

\begin{figure}
\includegraphics[width=3.4in]{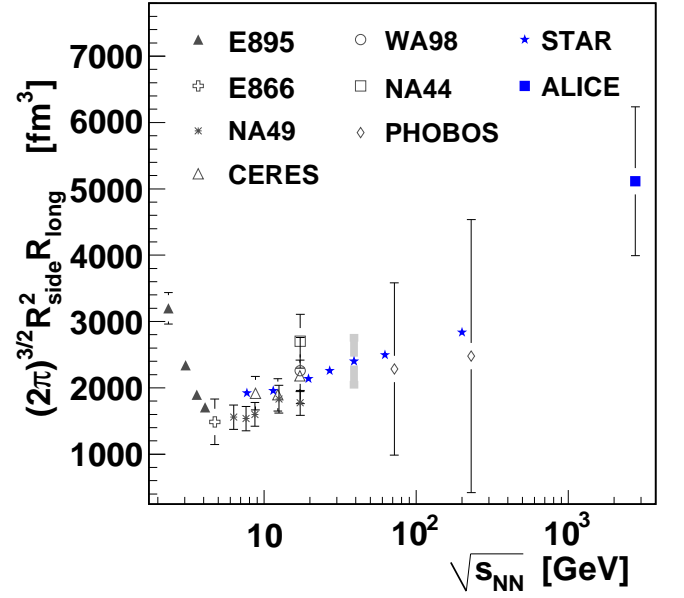} 
\caption{\label{F9b}  (Color online) The beam energy dependence of the volume, 
$V=(2\pi)^{3/2}R^{2}_{\mathrm{side}}R_{\mathrm{long}}$, of the regions of 
homogeneity at kinetic freeze-out in central Au+Au, Pb+Pb and Pb+Au collisions 
with $\langle k_{T}\rangle\approx0.22$ GeV/$c$~\cite{RvsRootSPHOBOS,RvsRootSCERES,RvsRootSE895,RvsRootSNA44,RvsRootSNA49,RvsRootSE866,RvsRootSWA98,AliceHbtPLB2010}.  
The systematic errors for STAR points at all energies (from Table~\ref{T2}) are 
of similar size to error bar for 39 GeV, shown as a representative example.  
Errors on other results are statistical only, to emphasize the trend.  The PHOBOS 
points are offset 
in $\sqrt{s_{NN}}$ for clarity.  The text contains some discussion about 
variations in centrality, $\langle k_{T}\rangle$, and analysis techniques 
between different experiments.}
\end{figure}

The systematic errors for STAR points 
at all energies (from Table~\ref{T2}) are of similar size to error bar for 39 GeV, 
shown as a representative example.  Errors on other results are statistical only
to emphasize the trend.  

  Figure~\ref{F8} shows the $\langle m_{T} \rangle$ dependence of the HBT parameters for each 
energy.  As mentioned earlier, the decrease in transverse and longitudinal radii 
at higher $m_{T}$ are attributed to transverse and longitudinal flow
\cite{LisaRetiere,SinyukovZPhysC}.  Larger $m_{T}$ pairs are emitted from smaller emission regions 
with less correspondence to the size of the entire fireball.  For both 
$R_{\mathrm{out}}$ and $R_{\mathrm{side}}$ the different beam energies show similar 
trends both in magnitude and slope.  For $R_{\mathrm{long}}$, the slopes appear to 
remain similar 
for the different energies, but the magnitude of $R_{\mathrm{long}}$ increases
with energy for all centralities.  
From these observations, 
and considering Fig.~\ref{F7} showed the beam energy dependence for a single 
$k_{T}$ and centrality bin, it is apparent that similar dependencies on 
$\sqrt{s_{NN}}$ exist for all the studied centrality and $k_{T}$ ranges.

  The multiplicity dependence of the HBT radii are presented in Fig.~\ref{F9} 
for two $k_{T}$ ranges with $\langle k_{T}\rangle\approx0.22$ GeV/$c$ and $\langle k_{T}\rangle\approx0.39$ 
GeV/$c$.  A few earlier measurements 
with similar $\langle k_{T}\rangle$ are shown as well.  It 
was observed in \cite{CuCuAuAu} that $R_{\mathrm{side}}$ and $R_{\mathrm{long}}$ both
follow a common universal trend at 62.4 and 200 GeV independent of the collision
species.  ALICE has recently shown p+p collisions exhibit a different 
multiplicity dependence with a smaller slope 
\cite{AliceHbtAdamsJPhysG2011,AliceHbtGramlingAIPConfProc2012}.  The difference
may be due to the interactions in the bulk medium formed in heavy ion 
collisions.

  The results from ALICE are at different $\langle k_{T} \rangle$ values.  To
get a similar $\langle k_{T} \rangle\approx 0.39$ GeV/$c$ estimate, the ALICE 
data points \cite{AliceHbtPLB2010} reported for 
$\langle k_{T} \rangle\approx 0.35$ GeV/$c$ and 
$\langle k_{T} \rangle\approx 0.44$ GeV/$c$ are averaged and plotted on 
Fig.~\ref{F9}.  There is some ambiguity in this approach as the different pair 
statistics at different $k_{T}$ are not accounted for when averaging this way.  
As demonstrated in \cite{AliceHbtPLB2010,AliceHbtAdamsJPhysG2011,AliceHbtGramlingAIPConfProc2012},
the universal trends for $R_{\mathrm{side}}$ and $R_{\mathrm{long}}$ continue
up to LHC energies.

When comparing different datasets from previous analyses 
\cite{RvsRootSCERES,AliceHbtPLB2010,RvsRootSE802}, there is an
uncertainty on the centrality caused by the different techniques that were
used to compute the average charged track multiplicity $\langle dN_{\mathrm{ch}}/d\eta \rangle$.
In this analysis, the standard STAR centrality definition was used at all
energies, where $\langle dN_{\mathrm{ch}}/d\eta \rangle$ is computed using
all events that pass the event selection cuts.
However,
it should be noted that this is an uncorrected value of 
$\langle dN_{\mathrm{ch}}/d\eta \rangle$ that underestimates the true value,
thus allowing for a qualitative comparison only with other experiments.  

\begin{figure}
\includegraphics[width=3.4in]{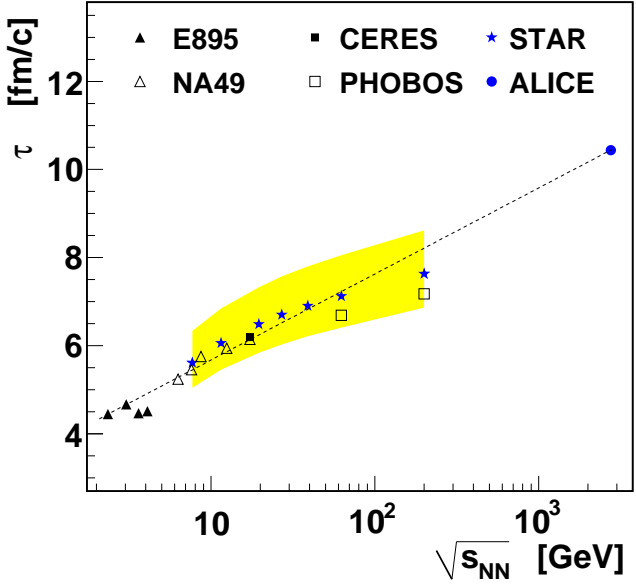} 
\caption{\label{F9c}  (Color online) The lifetime, $\tau$, of the system as a function of beam
 energy for central Au+Au collisions assuming a temperature of $T=0.12$ GeV at kinetic
 freeze-out.  The yellow band demonstrates the effect on $\tau$ of varying the assumed 
 temperature by $\pm0.02$ GeV.  Statistical uncertainties from the fits are smaller 
 than the data points.  The line extrapolates between the lowest and highest energy.  
 The text contains some discussion about variations in centrality and 
 analysis techniques between different experiments.} 
\end{figure}

  An estimate of the volume of the homogeneity regions, 
$V=\left( 2\pi\right)^{3/2}R^{2}_{\mathrm{side}}R_{\mathrm{long}}$, can be 
computed using the data in Fig.~\ref{F7}.  These values are plotted in 
Fig.~\ref{F9b} as a function of $\sqrt{s_{NN}}$.  The STAR results are all for 
$0$-$5\%$ central collisions with $\langle k_{T}\rangle \approx 0.22$ GeV/$c$.  
Since the values are computed using the data in Fig.~\ref{F7}, all the same 
variations in centrality ranges and $\langle k_{T}\rangle$ values are present 
in the volume estimates too.  Earlier results from other experiments 
suggest a minimum between 
AGS and SPS energies.  The STAR results show a noticeable increase in volume at 
the higher energies while the 7.7 and 11.5 GeV points are almost the same, 
consistent with a minimum in the vicinity of 7.7 GeV. The ALICE point rises 
even further 
suggesting the regions of homogeneity 
are significantly larger in collisions at the LHC.

  The CERES collaboration \cite{CeresMeanFreePath} has found that a constant 
mean free path at freeze-out, 
\begin{equation}\label{E18}
\lambda_{F} \approx \frac{V}{\left(N_{\pi}\sigma_{\pi\pi}+N_{N}\sigma_{\pi N}\right)} \approx 1 fm, 
\end{equation}
leads naturally to a minimum in the energy dependence of the volume that is 
observed, assuming that the cross sections $\sigma_{\pi\pi}$ and $\sigma_{\pi N}$
depend weakly on energy, since the yields of pions and nucleons, $N_{\pi}$ and $N_{N}$, change with 
energy.  
Above 19.6 GeV, the ratio of 
$N_{\pi}\sigma_{\pi\pi}/N_{N}\sigma_{\pi N}$ remains rather constant and the 
denominator in Eq.~\ref{E18} increases with energy similar to the volume.  
Below 11.5 GeV, the $N_{N}\sigma_{\pi N}$ term becomes the dominant term and it
increases at lower energies as does the volume.
At higher energies, this scenario 
is consistent with the nearly universal trend of the volume on
$\langle dN_{\mathrm{ch}}/d\eta\rangle$ and, therefore, $R_{\mathrm{side}}$ and
$R_{\mathrm{long}}$ on 
$\langle dN_{\mathrm{ch}}/d\eta\rangle^{1/3}$~\cite{CuCuAuAu}.
It is interesting that the multiplicity dependence for 
$R_{\mathrm{side}}$ begins to deviate slightly from
this trend for 7.7 and 11.5 GeV in Fig.~\ref{F9} which is the same
region where the system changes from $\pi$-$N$ to $\pi$-$\pi$ dominant.  Also,
the argument above neglects the influence from less abundant species including 
kaons, but it has been observed that strangeness enhancement occurs in this
same region of $\sqrt{s_{NN}}$ ~\cite{NA49StrangenessEnhancement}.  

  Another change that occurs in this region is the rapid increase of $v_{2}$
around $\sqrt{s_{NN}}=$ 2-7 GeV.  In the region around 7.7 to 11.5 GeV, the slope of
$v_{2}\left(\sqrt{s_{NN}}\right)$ begins to level off
\cite{ALICEflowVsRootS,STARv2hadrons}.  A possibility is 
that the deviation of $R_{\mathrm{side}}$ for 7.7 and 11.5 GeV is related to 
the onset of flow induced space-momentum correlations.  The E802 results at 4.8
and 5.4 GeV in the right column of Fig.~\ref{F9} are qualitatively similar to
the STAR 7.7 GeV results for $R_{\mathrm{side}}$, but considering the STAR
$\langle dN_{\mathrm{ch}}/d\eta\rangle^{1/3}$ values are slightly
underestimated,
the E802 results probably deviate slightly more relative to the higher energies
than even the 7.7 GeV data.  For $R_{\mathrm{out}}$, on the other hand, the 
E802 results are significantly larger than the STAR 7.7 GeV points.  This 
could be consistent with the effects of flow.  Transverse flow should reduce 
the size of the regions of homogeneity and is expected to affect 
$R_{\mathrm{out}}$ much more than $R_{\mathrm{side}}$.  This was reflected 
already in the larger slope for the $\langle m_{T}\rangle$ dependence of 
$R_{\mathrm{out}}$ relative to $R_{\mathrm{side}}$ in 
Fig.~\ref{F8}.  
It would 
be interesting to study these trends at lower energies with a single detector
where many interesting physical changes are occuring simultaneously.

   An alternative explanation of the minimum observed in the volume measurement
in Fig.~\ref{F9b} is provided by Ultra-relativistic Quantum Molecular Dynamics
(UrQMD) calculations.  In \cite{urqmdVolumeRef}, UrQMD
also finds a minimum between AGS and SPS energies but, in this case, the cause
is related to a different type of change in the particle production mechanism.
At the lowest energies pions are produced by resonances, but as the energy
increases more pions are produced by color string fragmentation (accounting
for color degrees of freedom) which freeze-out
at an earlier, smaller stage (thus a smaller volume is measured).
At even higher energies, the large increase in pion 
yields cause the volume to increase once more. 
This explanation suggests that a change
from hadronic to partonic degrees of freedom cause the minimum in the volume
measurement. 
Allowing a mean field potential to act 
on these pre-formed hadrons (the color string fragments) 
leads UrQMD to predict $R_{\mathrm{out}}/R_{\mathrm{side}}$ values near the 
observed values ($\approx 1$) for the whole energy range from AGS to SPS 
\cite{urqmdPRB659_525_2008}.  Simultaneously, inclusion of the mean field for 
pre-formed hadrons causes UrQMD to reproduce the net proton rapidity 
distribution and slightly improves its prediction for $v_{2}(p_{T})$ at 
intermediate $p_{T}$.

  As one last application of the data, the lifetime of the collisions is 
extracted in a study analogous to Ref.~\cite{AliceHbtPLB2010}.  We also assume a 
kinetic freeze-out temperature of $T=0.12$ GeV and fit the data in Fig.~\ref{F8} 
using Eq.~\ref{E17}.  The results are plotted in Fig.~\ref{F9c}.  The STAR 
results are all for $0$-$5\%$ collisions with 
$\langle k_{T}\rangle \approx 0.22$ GeV/$c$.  Again, there are some variations
in the centrality ranges, as in Fig.~\ref{F7}, for the historical data.  The extracted
lifetime appears to increase from around 4.5 fm/$c$ at the lowest energies to 
around 7.5 fm/$c$ at 200 GeV, an increase of an approximate factor of 1.7. 
The ALICE point suggests a much longer lived system, above 
the trend observed at lower energies.  
Varying the temperature assumed in the fits to $T=0.10$ GeV to $T=0.14$ GeV causes 
the lifetimes to increase by $13\%$ and decrease by $10\%$, respectively, for 
all energies, as indicated by the yellow band.  As noted in 
\cite{AliceHbtPLB2010}, due to effects from non-zero 
transverse flow and chemical potential for pions, the use of Eq.~\ref{E17} may 
significantly underestimate the actual lifetimes.

\begin{figure}
\includegraphics[width=3.4in]{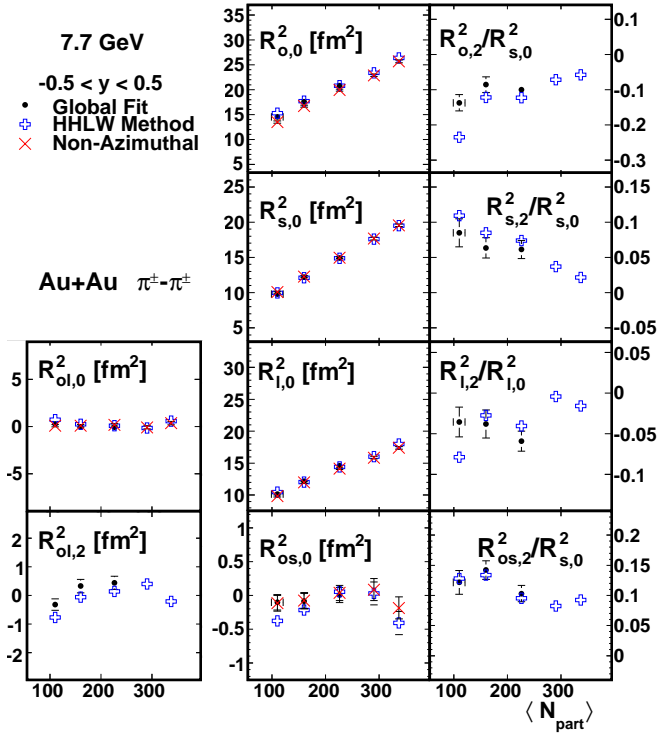} 
\caption{\label{F10}  (Color online) Centrality dependence of the Fourier coefficients that describe azimuthal oscillations of the HBT radii, at mid-rapidity 
($-0.5 < y < 0.5$), in 7.7 GeV collisions with $\langle k_{T}\rangle\approx$ 0.31 GeV/$c$.  Open symbols are results using separate Gaussian fits to each angular bin, the HHLW method.  Solid circles represent results using a single global fit to all
angular bins to directly extract the Fourier coefficients.  Crosses directly compare the azimuthally integrated radii and the
$0^{\mathrm{th}}$-order Fourier coefficients.  
Error bars include only statistical uncertainties.  The $0$-$5\%$ and $5$-$10\%$ global fit points have been excluded. }
\end{figure}

\begin{figure}
\includegraphics[width=3.4in]{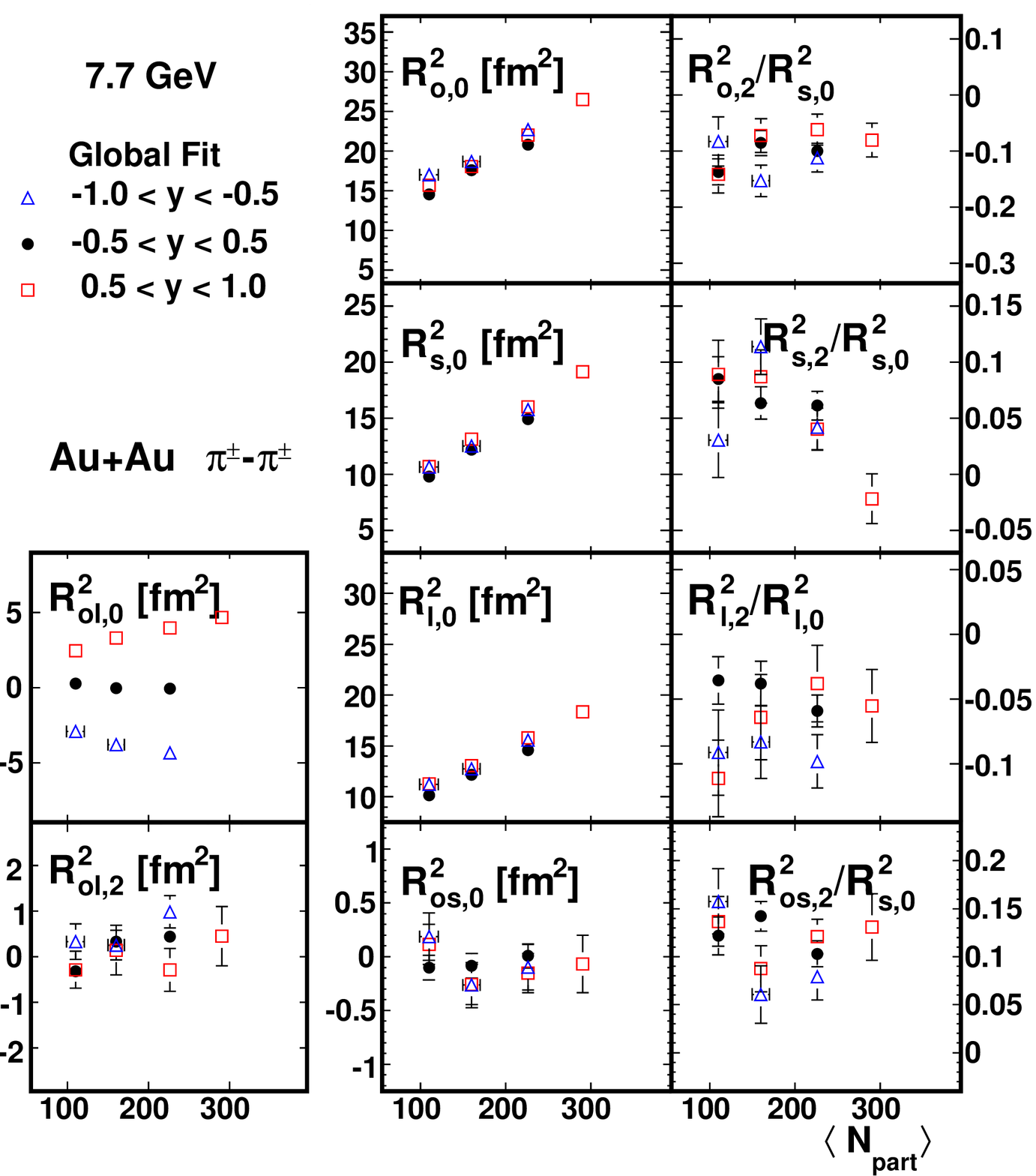} 
\caption{\label{F11}  (Color online) Centrality dependence of the Fourier coefficients that describe azimuthal oscillations of the HBT radii, at backward ($-1 < y
< -0.5$), forward ($0.5 < y < 1$) and mid ($-0.5 < y < 0.5$) rapidity, in 7.7 GeV collisions with $\langle k_{T}\rangle\approx$ 0.31 GeV/$c$ using the global fit method.  
Error bars include only statistical uncertainties.  The $0$-$5\%$ and two $5$-$10\%$ points have been excluded.}
\end{figure}

\begin{figure}
\includegraphics[width=3.4in]{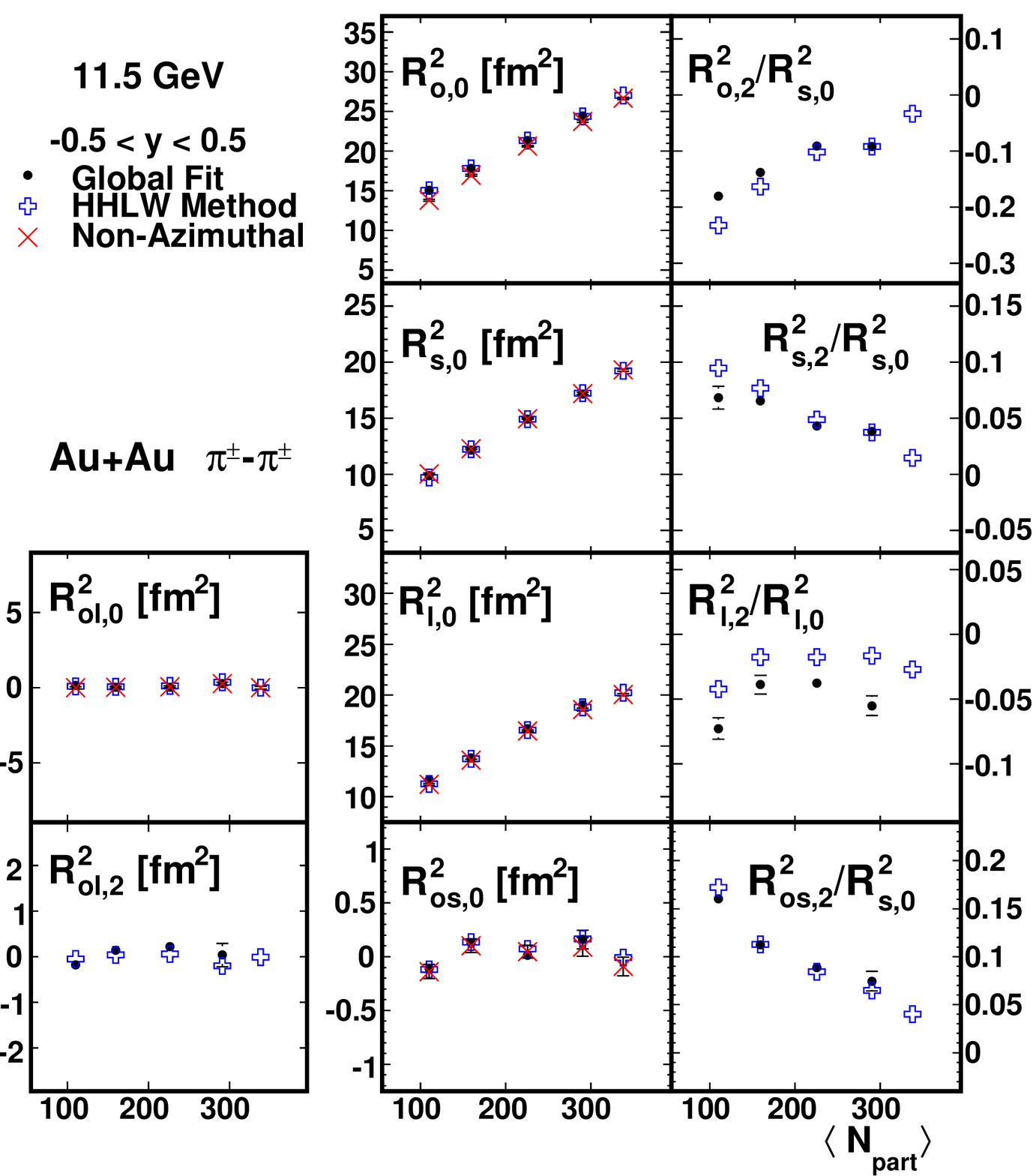} 
\caption{\label{F12}  (Color online) Centrality dependence of the Fourier coefficients that describe azimuthal oscillations of the HBT radii, at mid-rapidity 
($-0.5 < y < 0.5$), in 11.5 GeV collisions with $\langle k_{T}\rangle\approx$ 0.31 GeV/$c$.  
The symbols have the same meaning as in Fig.~\ref{F10}.  Error bars include only statistical uncertainties.  The $0$-$5\%$ global fit point is excluded.}
\end{figure}

\begin{figure}
\includegraphics[width=3.4in]{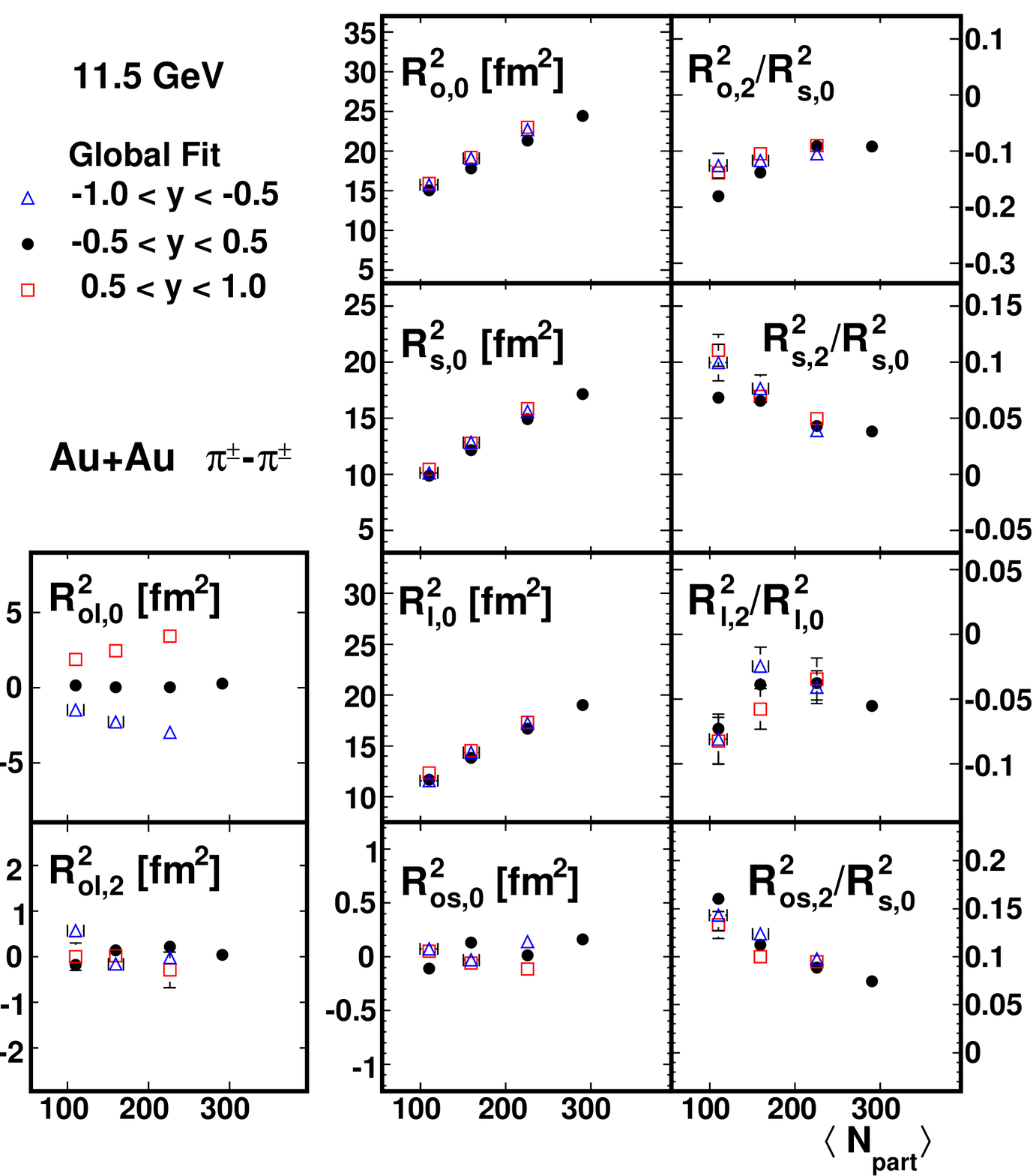} 
\caption{\label{F13}  (Color online) Centrality dependence of the Fourier coefficients that describe azimuthal oscillations of the HBT radii, at backward ($-1 < y
< -0.5$), forward ($0.5 < y < 1$) and mid ($-0.5 < y < 0.5$) rapidity, in 11.5 GeV collisions with $\langle k_{T}\rangle\approx$ 0.31 GeV/$c$ using the global fit method.  
Error bars include only statistical uncertainties.  The $0$-$5\%$ and two $5$-$10\%$ points have been excluded.}
\end{figure}

\begin{figure}
\includegraphics[width=3.4in]{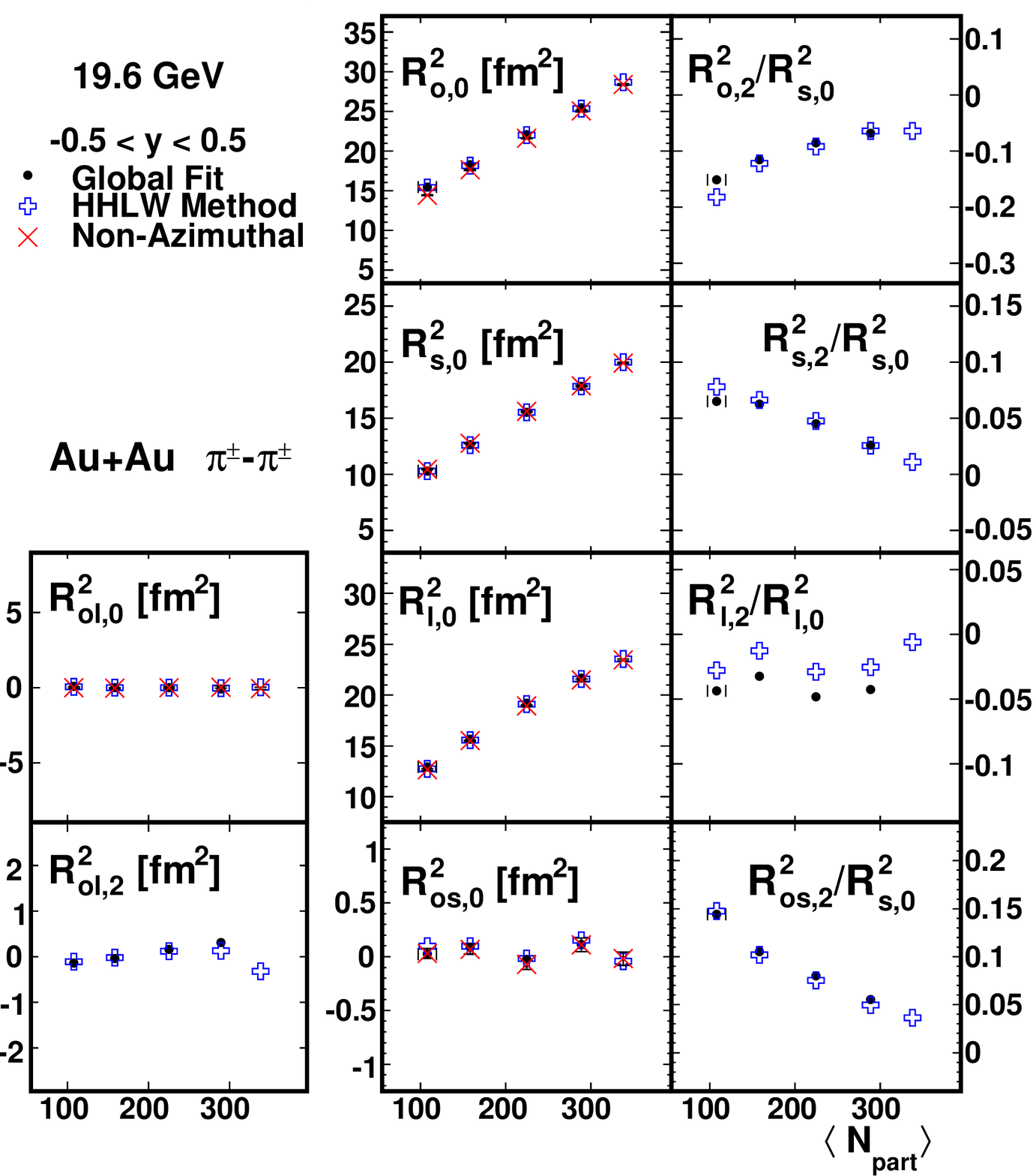}
\caption{\label{F14}  (Color online) Centrality dependence of the Fourier coefficients that describe azimuthal oscillations of the HBT radii, at mid-rapidity ($-0.5 < y <
0.5$), in 19.6 GeV collisions with $\langle k_{T}\rangle\approx$ 0.31 GeV/$c$.
The symbols have the same meaning as in Fig.~\ref{F10}.  Error bars include only statistical uncertainties.  The $0$-$5\%$ global fit point is excluded.}
\end{figure}

\begin{figure}
\includegraphics[width=3.4in]{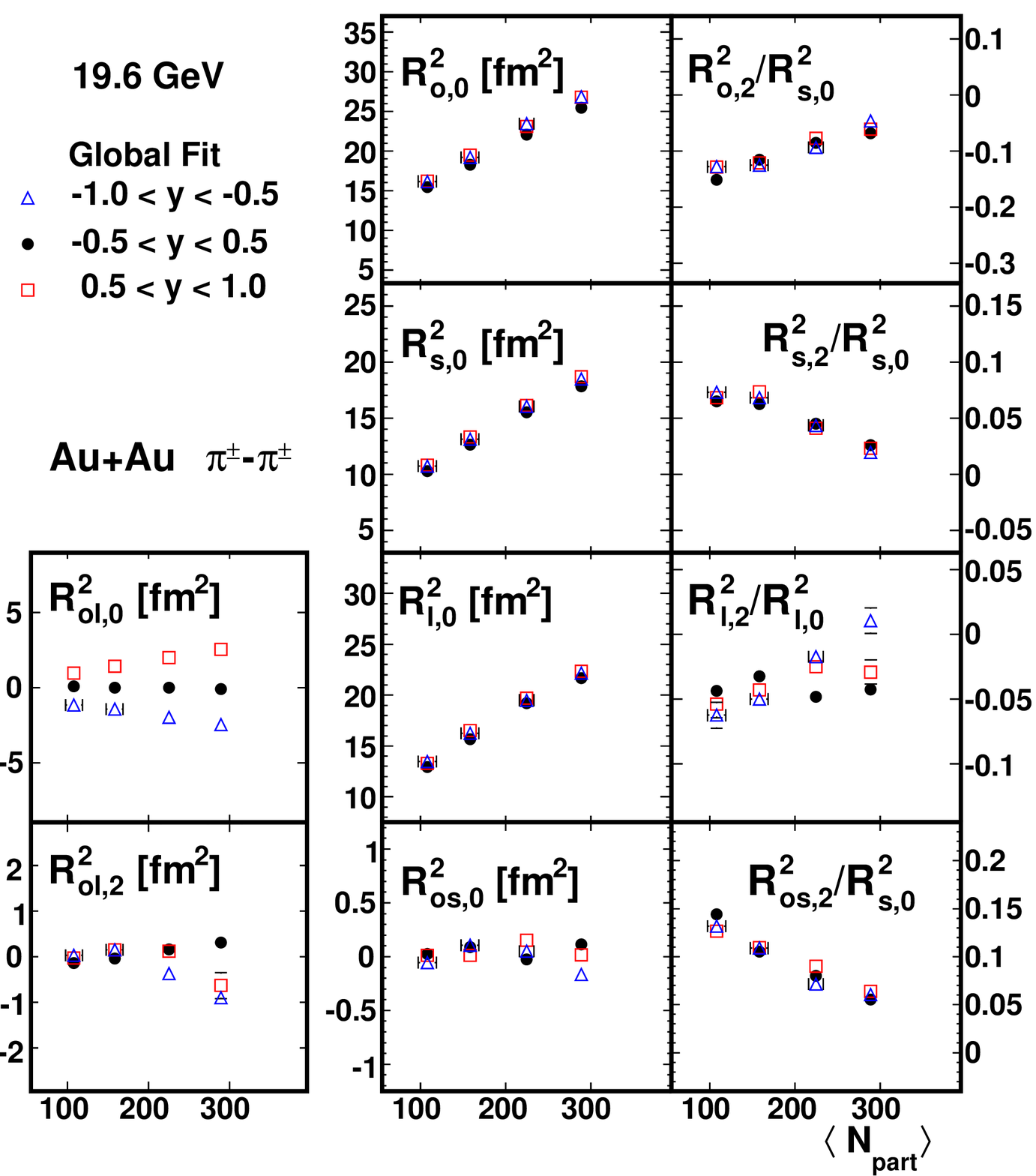} 
\caption{\label{F15}  (Color online) Centrality dependence of the Fourier coefficients that describe azimuthal oscillations of the HBT radii, at backward ($-1 < y
< -0.5$), forward ($0.5 < y < 1$) and mid ($-0.5 < y < 0.5$) rapidity, in 19.6 GeV collisions with $\langle k_{T}\rangle\approx$ 0.31 GeV/$c$ using the global fit method.  
Error bars include only statistical uncertainties.  The $0$-$5\%$ global fit point is excluded.}
\end{figure}

\subsection{Azimuthally differential HBT}\label{S5b}

\begin{figure}
\includegraphics[width=3.4in]{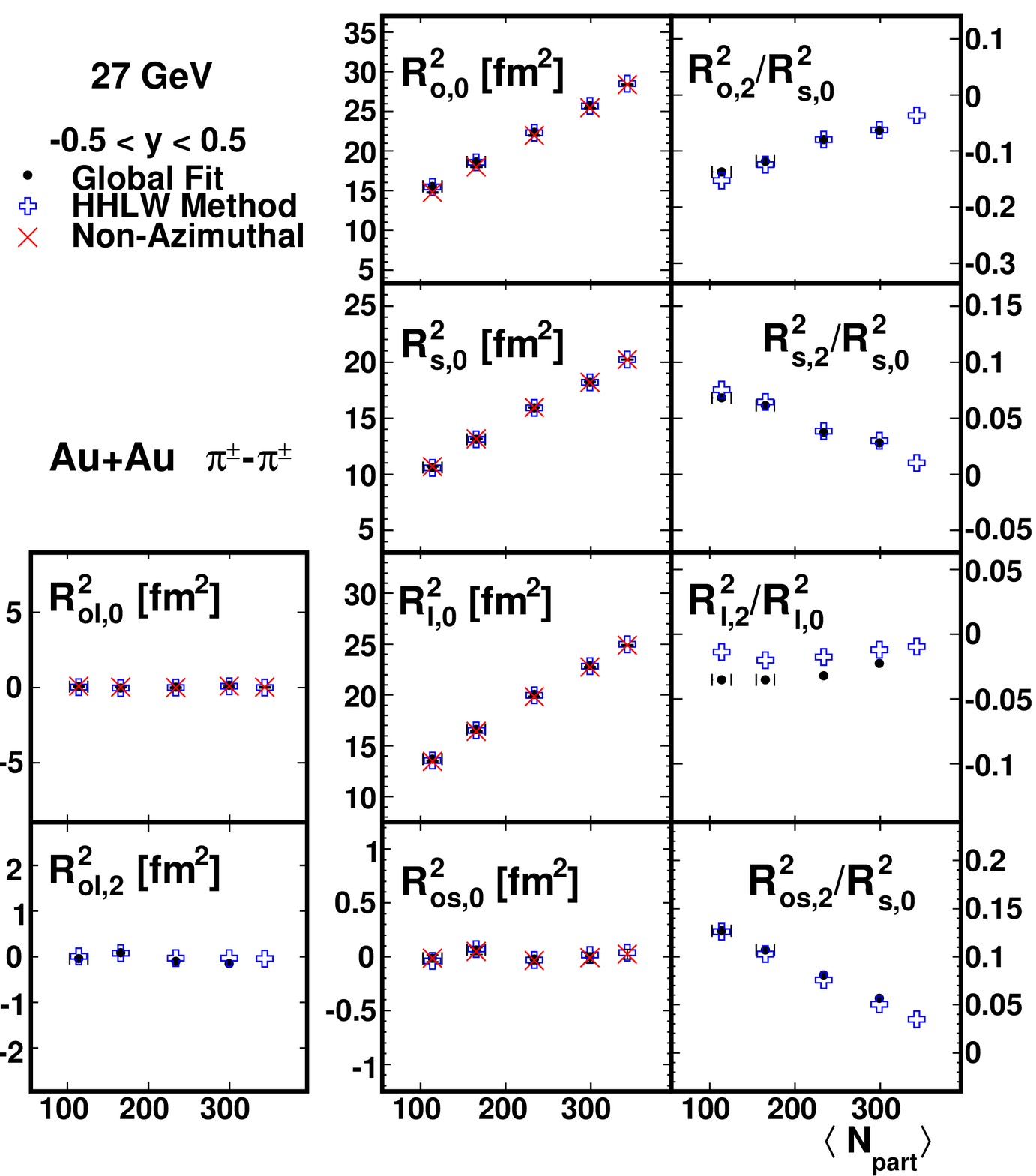} 
\caption{\label{F16}  (Color online) Centrality dependence of the Fourier coefficients that describe azimuthal oscillations of the HBT radii, at mid-rapidity ($-0.5 < y <
0.5$), in 27 GeV collisions with $\langle k_{T}\rangle\approx$ 0.31 GeV/$c$.  
The symbols have the same meaning as in Fig.~\ref{F10}.  Error bars include only statistical uncertainties.  The $0$-$5\%$ global fit method point is excluded.}
\end{figure}
\begin{figure}
\includegraphics[width=3.4in]{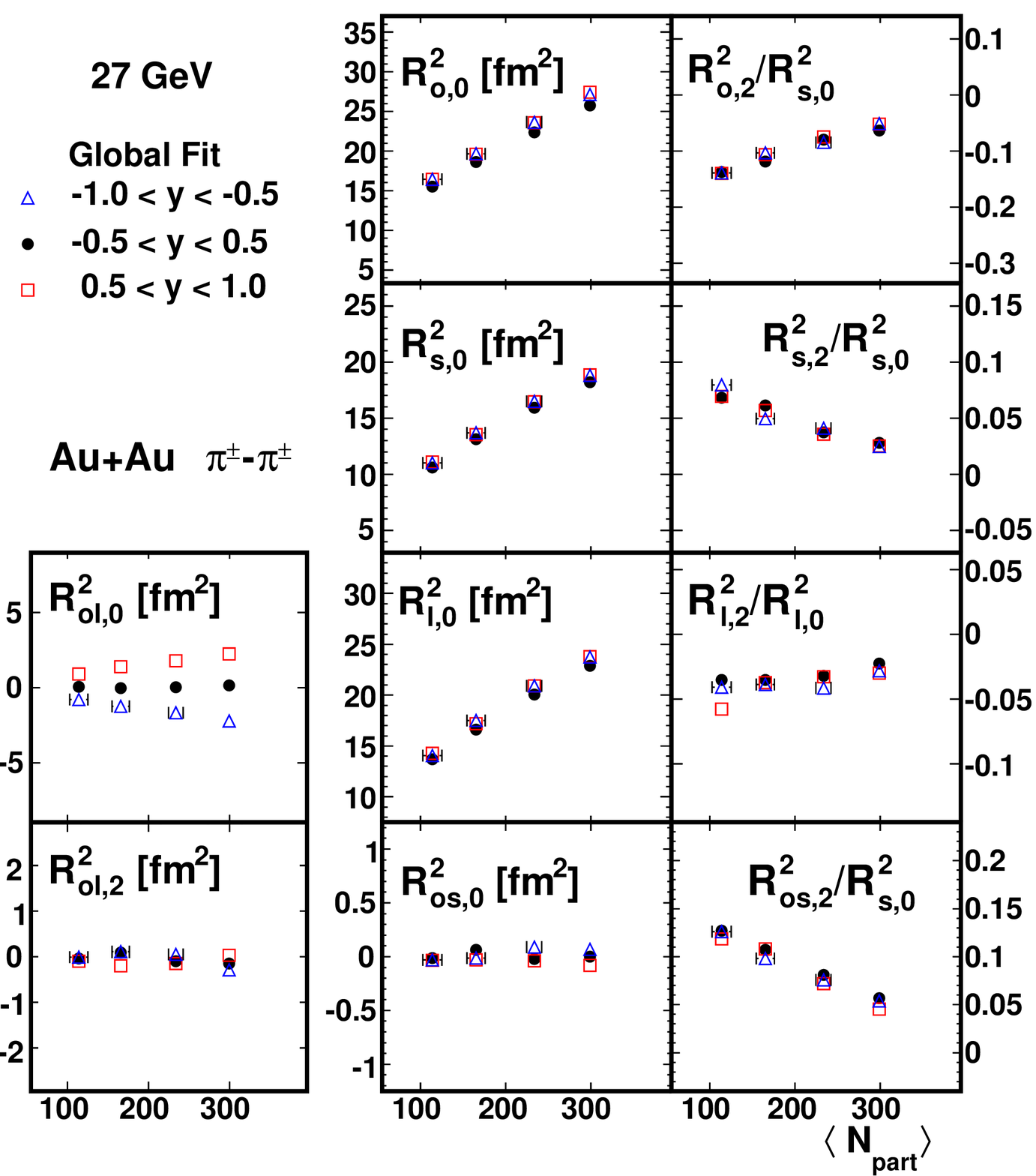} 
\caption{\label{F17}  (Color online) Centrality dependence of the Fourier coefficients that describe azimuthal oscillations of the HBT radii, at backward ($-1 < y
< -0.5$), forward ($0.5 < y < 1$) and mid ($-0.5 < y < 0.5$) rapidity, in 27 GeV collisions with $\langle k_{T}\rangle\approx$ 0.31 GeV/$c$ using the global fit method.
Error bars include only statistical uncertainties.  The $0$-$5\%$ global fit point is excluded.}
\end{figure}
\begin{figure}
\includegraphics[width=3.4in]{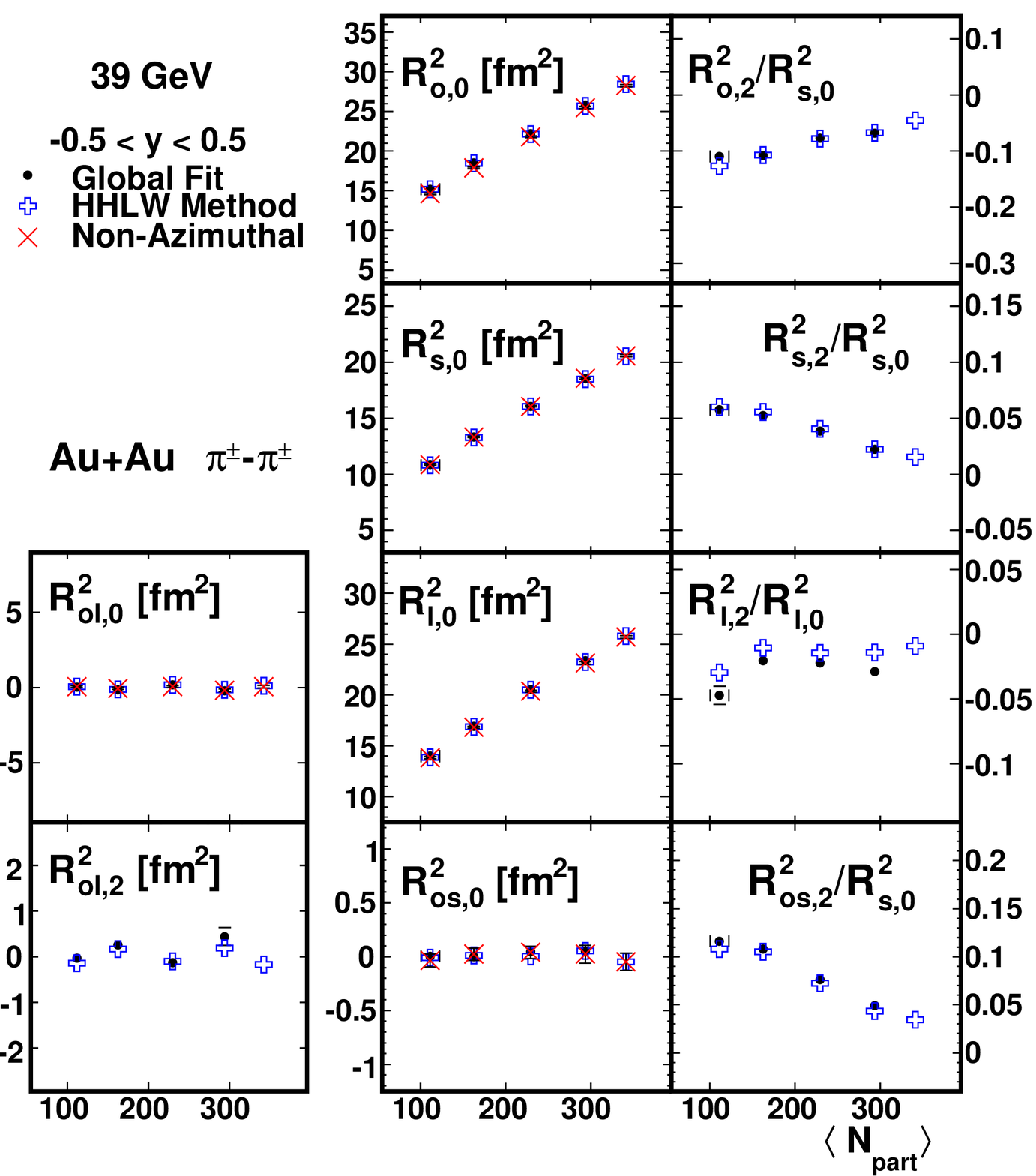} 
\caption{\label{F18}  (Color online) Centrality dependence of the Fourier coefficients that describe azimuthal oscillations of the HBT radii, at mid-rapidity ($-0.5 < y <
0.5$), in 39 GeV collisions with $\langle k_{T}\rangle\approx$ 0.31 GeV/$c$.  
The symbols have the same meaning as in Fig.~\ref{F10}.  Error bars include only statistical uncertainties.  The $0$-$5\%$ global fit method point is excluded.}
\end{figure}
\begin{figure}
\includegraphics[width=3.4in]{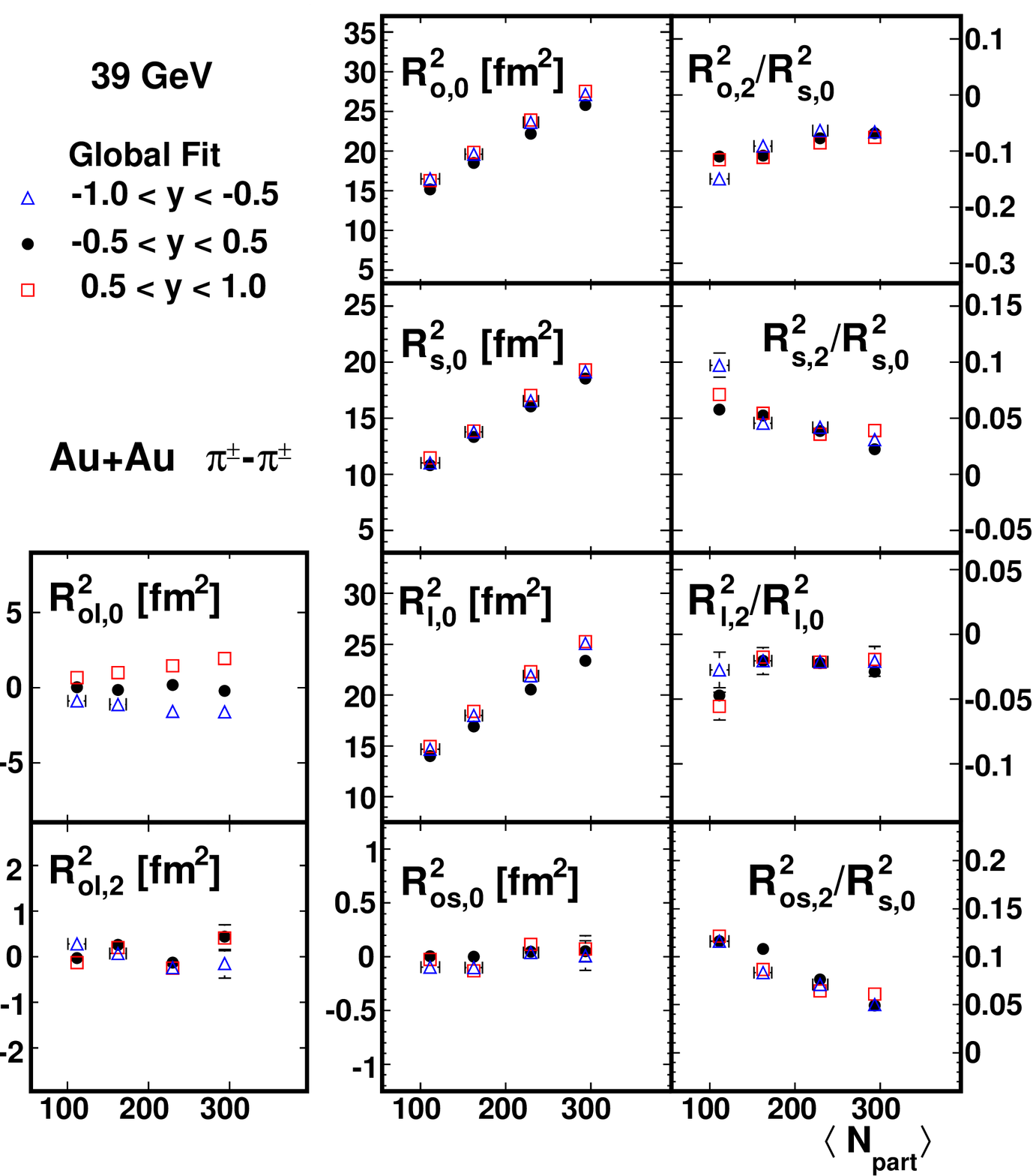} 
\caption{\label{F19}  (Color online) Centrality dependence of the Fourier coefficients that describe azimuthal oscillations of the HBT radii, at backward ($-1 < y
< -0.5$), forward ($0.5 < y < 1$) and mid ($-0.5 < y < 0.5$) rapidity, in 39 GeV collisions with $\langle k_{T}\rangle\approx$ 0.31 GeV/$c$ using the global fit method.
Error bars include only statistical uncertainties.  The $0$-$5\%$ global fit point is excluded.}
\end{figure}
\begin{figure}
\includegraphics[width=3.4in]{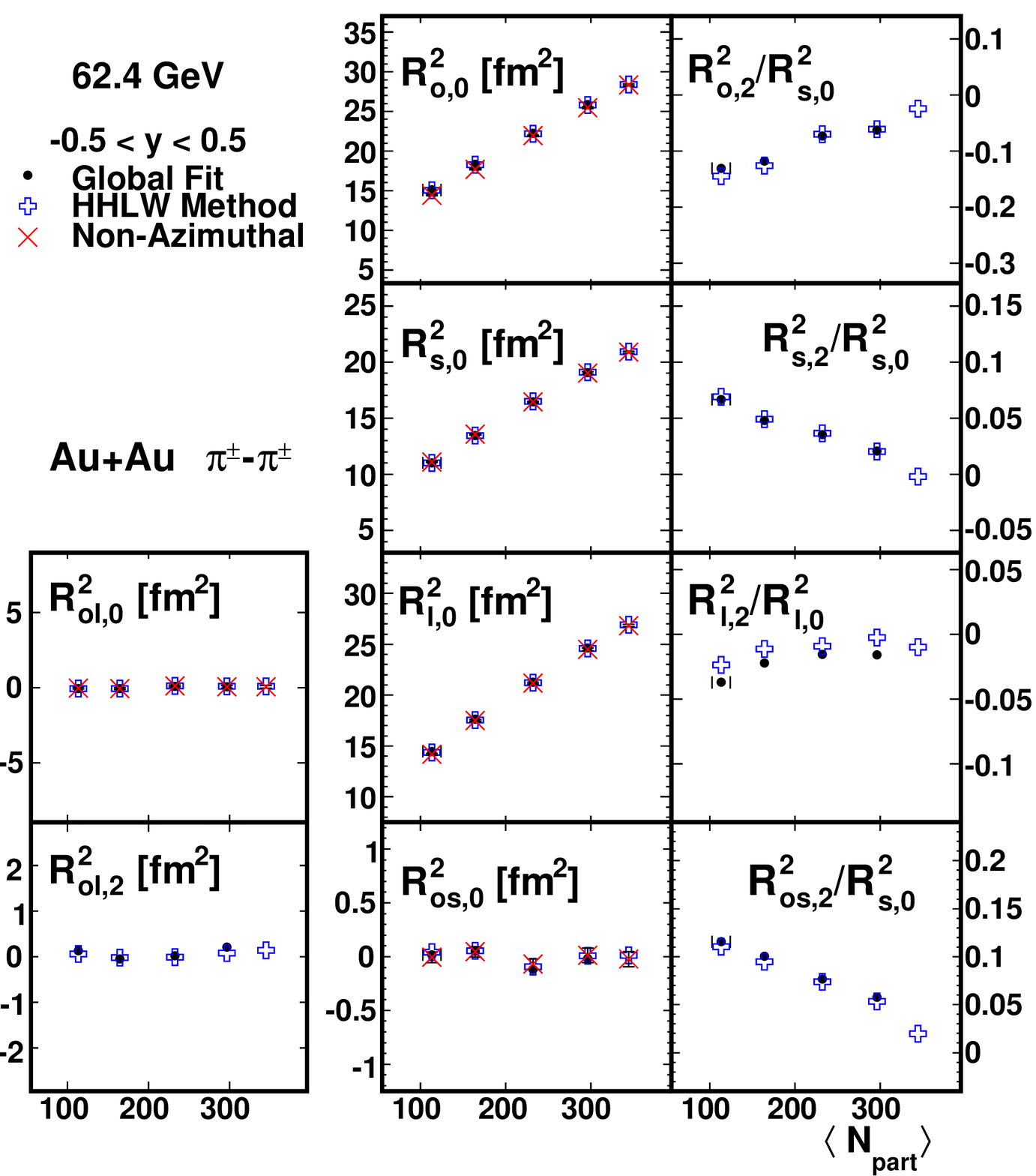} 
\caption{\label{F20}  (Color online) Centrality dependence of the Fourier coefficients that describe azimuthal oscillations of the HBT radii, at mid-rapidity ($-0.5 < y <
0.5$), in 62.4 GeV collisions with $\langle k_{T}\rangle\approx$ 0.31 GeV/$c$.  
The symbols have the same meaning as in Fig.~\ref{F10}.  Error bars include only statistical uncertainties.  The $0$-$5\%$ global fit method point is excluded.}
\end{figure}
\begin{figure}
\includegraphics[width=3.4in]{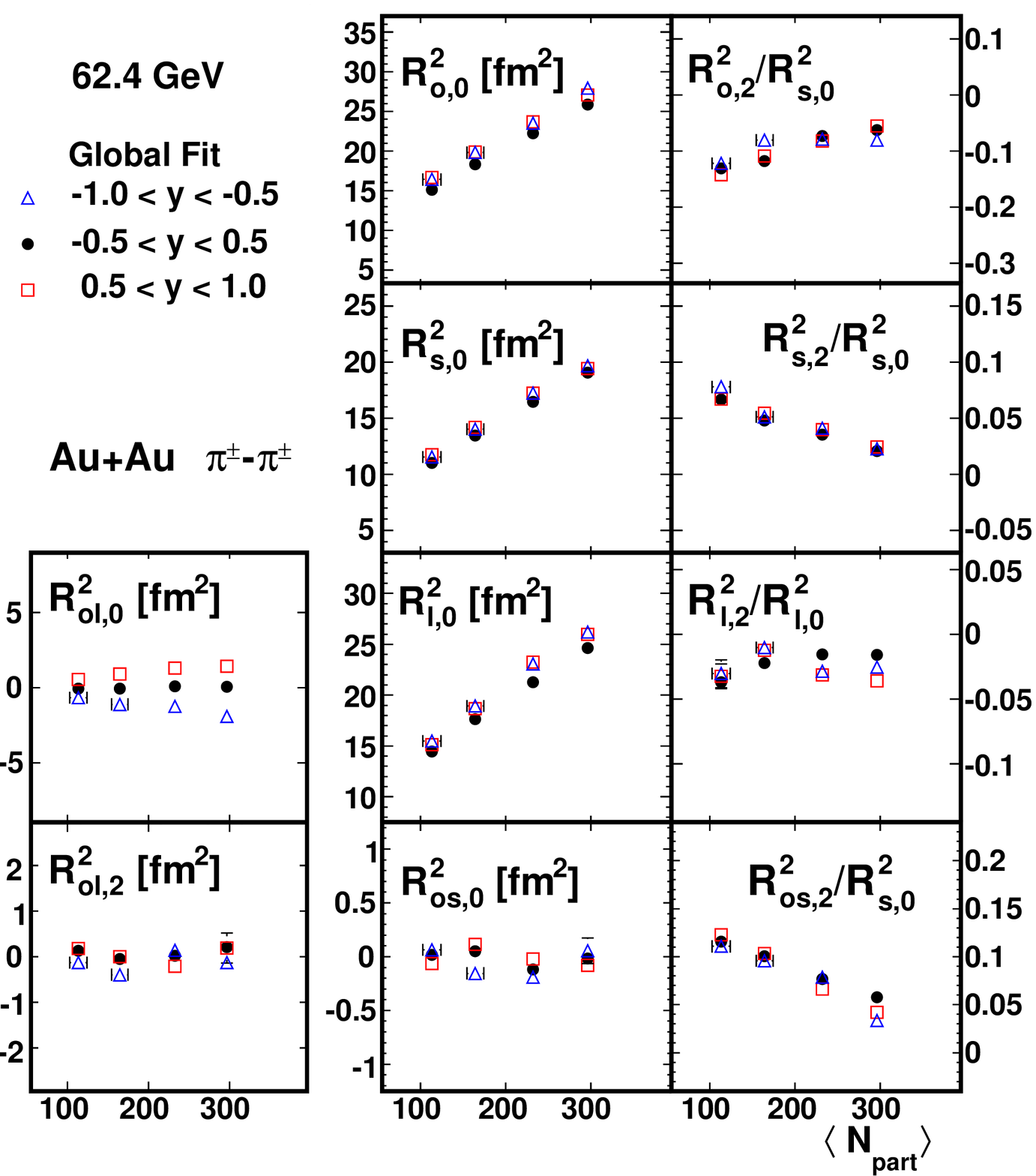} 
\caption{\label{F21}  (Color online) Centrality dependence of the Fourier coefficients that describe azimuthal oscillations of the HBT radii, at backward ($-1 < y
< -0.5$), forward ($0.5 < y < 1$) and mid ($-0.5 < y < 0.5$) rapidity, in 62.4 GeV collisions with $\langle k_{T}\rangle\approx$ 0.31 GeV/$c$ using the global fit method.
Error bars include only statistical uncertainties.  The $0$-$5\%$ global fit point is excluded.}
\end{figure}
\begin{figure}
\includegraphics[width=3.4in]{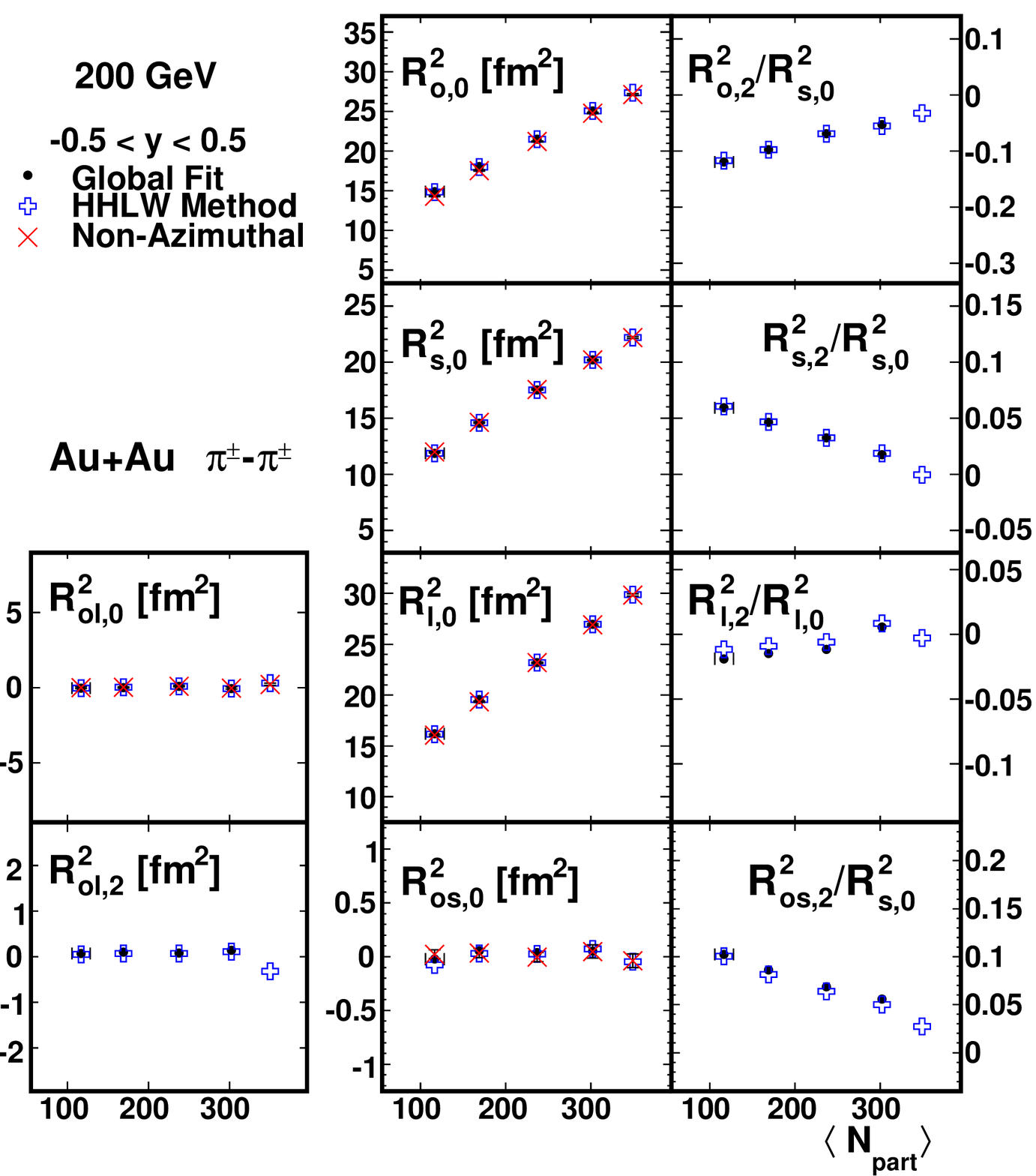} 
\caption{\label{F22}  (Color online) Centrality dependence of the Fourier coefficients that describe azimuthal oscillations of the HBT radii, at mid-rapidity ($-0.5 < y <
0.5$), in 200 GeV collisions with $\langle k_{T}\rangle\approx$ 0.31 GeV/$c$.  
The symbols have the same meaning as in Fig.~\ref{F10}.  Error bars include only statistical uncertainties.  The $0$-$5\%$ global fit method point is excluded.}
\end{figure}
\begin{figure}
\includegraphics[width=3.4in]{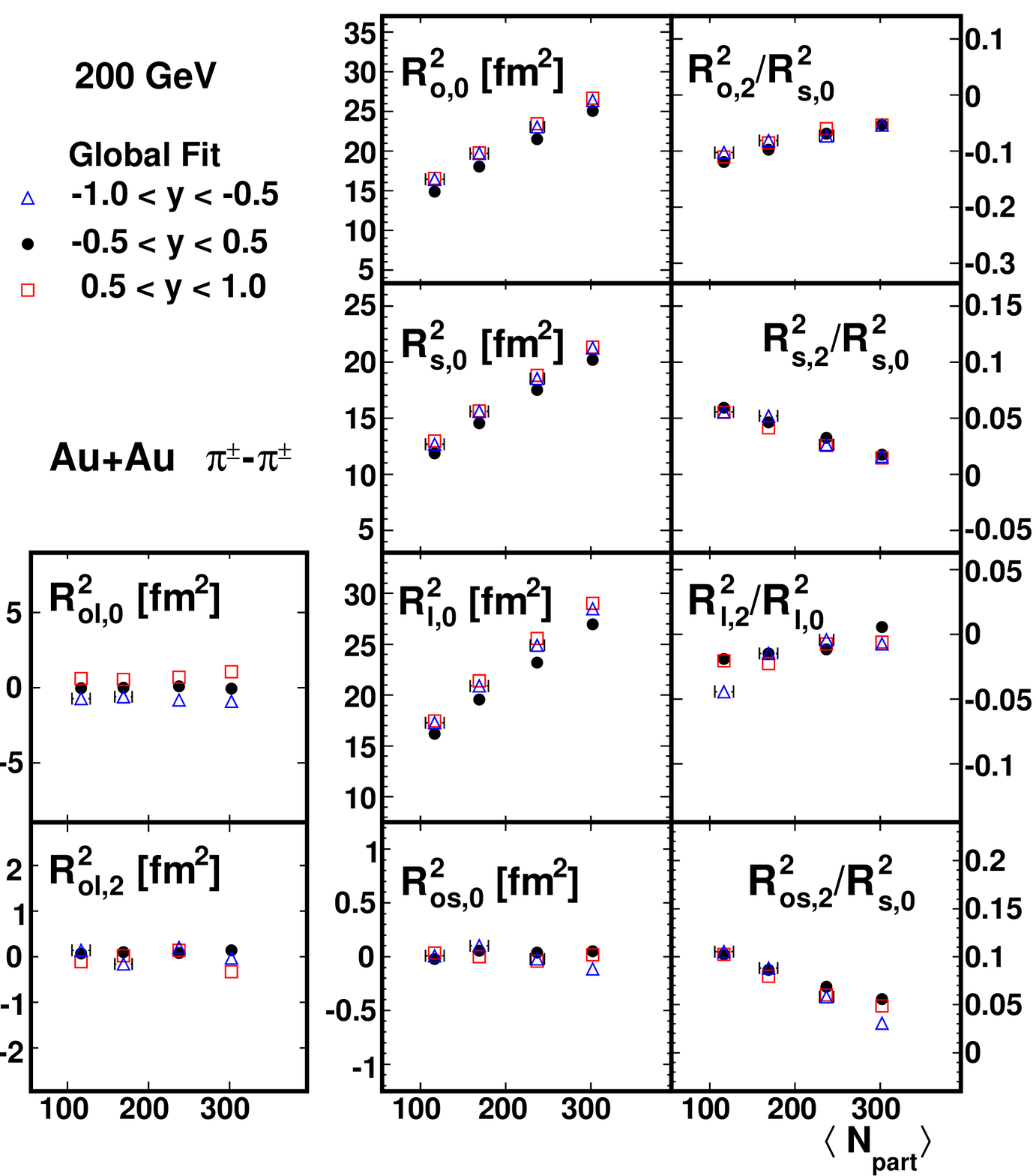} 
\caption{\label{F23}  (Color online) Centrality dependence of the Fourier coefficients that describe azimuthal oscillations of the HBT radii, at backward ($-1 < y
< -0.5$), forward ($0.5 < y < 1$) and mid ($-0.5 < y < 0.5$) rapidity, in 200 GeV collisions with $\langle k_{T}\rangle\approx$ 0.31 GeV/$c$ using the global fit method.
Error bars include only statistical uncertainties.  The $0$-$5\%$ global fit point is excluded.}
\end{figure}

  The detailed results of the azimuthally differential analysis are presented in
Figs.~\ref{F10} through~\ref{F24}.  Earlier, Fig.~\ref{F4} presented an example
of the second order oscillations of the HBT radii relative to the event plane for
a single energy, centrality, and rapidity.  These second order oscillations are
represented by $0^{\mathrm{th}}$- and $2^{\mathrm{nd}}$-order Fourier coefficients, 
as described in Sec.~\ref{Sb4a}.  The Fourier coefficients are presented as a 
function of $N_{\mathrm{part}}$ in two figures for each energy, starting with
Figs.~\ref{F10} and~\ref{F11} for 7.7 GeV and continuing through Figs.~\ref{F22}
and~\ref{F23} for 200 GeV.  For each energy, the first figure compares mid-rapidity
results from the HHLW and global fit methods while the second compares forward,
backward, and mid-rapidity results obtained using the global fit method.  Each
set of Fourier coefficients for a given $N_{\mathrm{part}}$ (centrality), rapidity,
and energy encodes all the information for oscillations similar to those in 
Fig.~\ref{F4}.

  In each of the figures showing the Fourier coefficients, the $0^{\mathrm{th}}$-order 
coefficients are presented in the middle column, for the squared radii in the 
out, side and long directions ($R^{2}_{o,0}$, $R^{2}_{s,0}$, $R^{2}_{l,0}$) and 
the out-side cross term ($R^{2}_{os,0}$).  These values are expected to
correspond to radii from the azimuthally integrated analysis.  This 
correspondence is demonstrated in the first Fourier coefficient figure for each
energy which also includes the azimuthally integrated results (red crosses) for
direct comparison.  As in the
azimuthally integrated case, the diagonal radii increase with centrality while
the $R^{2}_{os,0}$ cross term remains about zero for all centralities.  In the 
right column of these figures, ratios of $2^{\mathrm{nd}}$-order to 
$0^{\mathrm{th}}$-order coefficients are presented, also for the out, side, 
long and out-side parameters.  The ratios that are presented have been connected 
to the freeze-out geometry, especially for the $R^{2}_{s,2}/R^{2}_{s,0}$ case.  
The left column of each of the figures contains the parameters for the out-long 
cross term.  The $0^{\mathrm{th}}$-order values, $R^{2}_{ol,0}$, are non-zero 
away from mid-rapidity and show interesting dependence on energy and centrality 
that will be discussed later.

\subsubsection{Comparison of fit methods}\label{S5b1}

  This section provides a comparison of the HHLW fit method and the global fit 
method used in the azimuthally differential analysis at mid-rapidity.
The first Fourier coefficient figure for each energy is relevant for this
discussion.  For Sec.~\ref{S5b2}, the second Fourier coefficient figure for
each energy is relevant for the discussion of centrality and rapidity
dependence of the Fourier coefficients.

  The results using the two fit methods are generally consistent for most of the
parameters.  For each energy, the first figure compares the Fourier coefficients
from the two fit methods at mid-rapidity.  Forward and backward rapidity results 
are not included as some of the results become unreliable in a few cases.
The reason is that at the lowest energies statistics limits the reliability,
ofthe HHLW fit method, especially for 7.7 GeV which has
the fewest events and the lowest multiplicity per event.  The forward and
backward rapidity regions have even lower statistics due to the narrower
window of rapidity, $\Delta y=0.5$ rather than $\Delta y=1$.  As seen in 
Fig.~\ref{F3}, the event plane
resolutions are much lower at these energies as well which can amplify noise in 
the correlation function when the correction algorithm is applied.  The 
correction algorithm does not distinguish between a real signal and a 
statistical variation.  The amplitude of the variations is increased in either case.  The global 
fit method was designed to minimize this problem by only applying the 
correction algorithm to the denominator which has an order of magnitude higher 
statistics than the numerator.

  The $0^{\mathrm{th}}$-order Fourier coefficients are expected to be consistent 
with the radii in the azimuthally integrated analysis.  Therefore, the
$0^{\mathrm{th}}$-order, squared radii should increase smoothly with 
$N_{\mathrm{part}}$ (as in the middle column of 
Figs.~\ref{F10} through~\ref{F23}).
For the $0^{\mathrm{th}}$-order terms good agreement with the azimuthally integrated
results is observed for both the HHLW and global fit methods, except a
few cases at the lowest energies.  Especially for 7.7 GeV, with the HHLW
fit method, several points, primarily the most peripheral and more central
(lowest statistics and resolution) points, were found to deviate quite 
significantly from this
trend.  All of these points are excluded in the figures since they are unreliable.
In the same cases, however, the global fit method remains consistent with the
non-azimuthal radii.  Projections of the fits on the out, side, and long axes
show the HHLW fit method
results do not match well with the data in such cases.  In particular, the
$90^{\circ}$ bin suffers most from low statistics (fewer tracks are directed
out of the reaction plane) which affects both the $0^{\mathrm{th}}$- and 
$2^{\mathrm{nd}}$-order coefficients when each bin is fit separately.  The
global fit method results 
are 
somewhat more reliable in these low statistics and low resolution cases.

  As noted earlier, there is a difference in the oscillation amplitude
for the long direction, $R^{2}_{l,2}$, obtained from the two methods.  
This is shown clearly in Fig.~\ref{F4} where the global fit method extracts
a larger oscillation amplitude.  From the first Fourier coefficient figure
at each energy, the ratio $R^{2}_{l,2}/R^{2}_{l,0}$ is systematically further
below zero for the global fit method results.  This is a systematic difference,
independent of centrality and energy, related to the different parameterizations
in the two fit methods.

  For reasons discussed in Sec.~\ref{Sb4c}, results using the global fit method 
are not shown for the most central $0$-$5\%$ data, as well as a few $5$-$10\%$ 
cases for 7.7 and 11.5 GeV where the statistics are low.
Still, in all cases, the fit projections from the global fit 
method better match the data, there is better agreement between forward and 
backward as well as mid-rapidity results and, as discussed in Sec.~\ref{Sb4c}, 
the errors are not underestimated as they are for the HHLW fit method.  
Therefore, results using the global fit method are used later when discussing 
the freeze-out shape.

\subsubsection{Fourier components}\label{S5b2}

  The trends exhibited by the Fourier coefficients are qualitatively similar
for all energies.  The $0^{\mathrm{th}}$-order coefficients are consistent with the
non-azimuthal results.  Like in the non-azimuthal results, the increase of the
$0^{\mathrm{th}}$-order coefficients for more central data is related to
the increasing volume of the homogeneity regions in more central events. Since
the ratios of $2^{\mathrm{nd}}$- to $0^{\mathrm{th}}$-order results are related 
to the freeze-out shape, the trends are expected to extrapolate toward zero for 
more central, more round collisions.  The right column of the Fourier 
coefficient figures for each energy demonstrate that this behavior is observed.  
For each HBT radius, the ratios of $2^{\mathrm{nd}}$- to $0^{\mathrm{th}}$-order 
coefficients follow similar trends for all energies, rapidities, and 
centralities.  This means that the $2^{\mathrm{nd}}$-order coefficients (half
the oscillation amplitudes) have the same sign in all these cases.  Therefore, 
the data requires that all energies, rapidity ranges, and centralities must 
exhibit oscillations of the HBT radii that are qualitatively similar to those in 
Fig.~\ref{F4}.  The Fourier coefficients for all three rapidities are 
similar in most cases, especially in the $R^{2}_{s,2}/R^{2}_{s,0}$ values 
for $10$-$20\%$ and $20$-$30\%$ centralities used later in the excitation 
function for the freeze-out eccentricity.

  One interesting feature occurs in the $R^{2}_{ol,0}$ parameter at forward
and backward rapidity.  This parameter exhibits both centrality and energy 
dependence that may be relevant for constraining future model studies. 
The centrality dependence is shown in the upper panels in the left column of 
Figs.~\ref{F11},~\ref{F13},~\ref{F15},~\ref{F17},~\ref{F19},~\ref{F21}, and 
~\ref{F23}.  As discussed earlier, this term averages to zero for results 
centered at mid-rapidity, but is otherwise non-zero.  At the lowest energy,
the $R^{2}_{ol,0}$ offset is quite large (Fig.~\ref{F11}) and increases in a 
linear manner with $N_{\mathrm{part}}$.  At higher energies, 
although the linear trend with $N_{\mathrm{part}}$ 
remains, the slope decreases for larger $\sqrt{s_{NN}}$.  For the 200 GeV 
results in Fig.~\ref{F23}, the slope and values are quite small compared to the 
7.7 GeV case, for instance.  As discussed in 
Sec.~\ref{S3b}, this non-zero cross term corresponds to a tilt in the 
$q_{\mathrm{out}}$-$q_{\mathrm{long}}$ plane.  The non-zero value of the
cross term means there is a correlation between the relative momentum of
particle pairs in the out and long directions. 

  Two considerations affect how $R^{2}_{ol,0}$ (or any of the radii) are
related to physical parameters of interest.  One is the frame in which the 
correlation function is constructed (fixed center of mass, LCMS, etc.)
~\cite{RolCrossTermNix,BPcoordinatesRolcrossterm}.  The 
other involves the assumptions that enter a particular analytical model 
of the source distribution (static, longitudinal flow, transverse flow, 
boost-invariance, etc.) that is required to relate the extracted
fit parameters (radii) to physical quantities such as freeze-out duration or
total lifetime~\cite{RolCrossTermNix,BPcoordinatesRolcrossterm}.

    Assume for the moment that radii are measured in the LCMS frame, as in this
analysis.  In models with longitudinal expansion, breaking of boost-invariance
results in non-zero values of the $R^{2}_{ol,0}$ cross term away from
mid-rapidity~\cite{RolCrossTermNix,BPcoordinatesRolcrossterm}.  The reason is 
that the LCMS and local rest frame of the source only coincide in the 
boost-invariant model~\cite{RolCrossTermNix}.  This is one example of how 
changing the model assumptions leads to a different relationship between the 
radii (including $R^{2}_{ol,0}$) and physical parameters.

     Alternatively, if the same analytical model is assumed but the measurement
is performed in different frames, the dependence of the radii on the physical parameters
will also change.  
Ref.~\cite{RolCrossTermNix} demonstrates that, assuming boost invariant longitudinal
expansion, measurement in a fixed frame, the LCMS frame, and a
generalized Yano-Koonin frame lead to three different relationships between the
fit parameters (radii) and physical quantities.   In 
Ref.~\cite{BPcoordinatesRolcrossterm}, a similar analytical model leads to a quite 
complex dependence of $R^{2}_{ol,0}$ on various physical quantities in the 
center of mass frame.  However, the expression greatly simplifies in the LCMS 
frame, leaving $R^{2}_{ol,0}$ directly proportional to the freeze-out duration
and other parameters.

\begin{figure}
\includegraphics[width=3.4in]{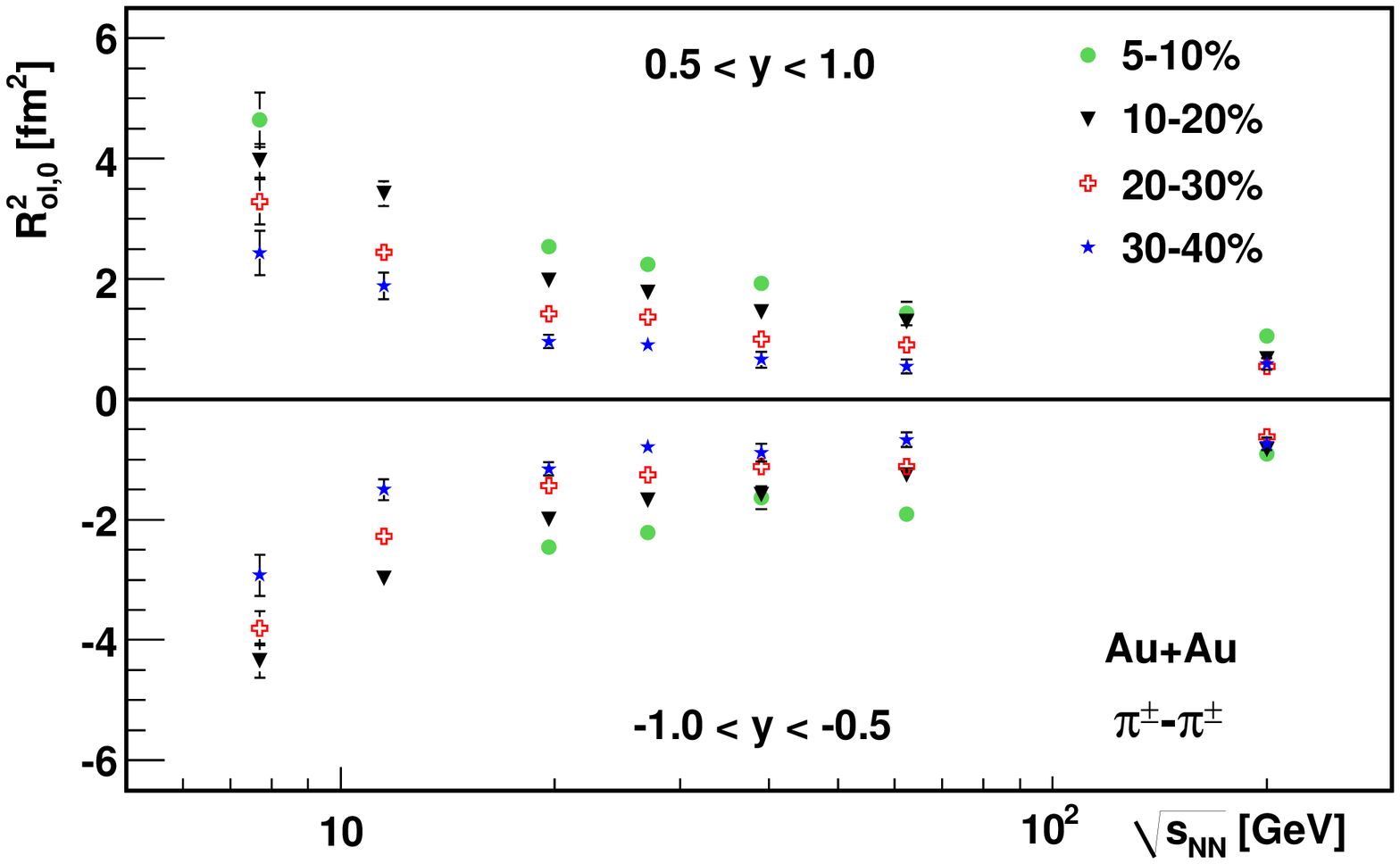} 
\caption{\label{F26}  (Color online) Beam energy dependence of the $R^{2}_{ol,0}$ cross term for forward and backward rapidity with $\langle k_{T}\rangle\approx$ 0.31 GeV/$c$.}
\end{figure}

  Fig.~\ref{F26} shows that, for each centrality, $R^{2}_{ol,0}$ decreases
smoothly toward zero at higher collision energy. 
It has been suggested \cite{BPcoordinatesRolcrossterm,RolCrossTermNix} that the 
quantity $R^2_{\mathrm{out}}-R^{2}_{\mathrm{side}}$ is sensitive to the 
duration of particle emission, $\Delta\tau$, 
which provided the main motivation for the past studies of 
$R_{\mathrm{out}}/R_{\mathrm{side}}$, summarized
in Fig.~\ref{F7}.  The $R^{2}_{ol}$ offset has also been associated with the 
duration of freeze-out and other parameters in a mathematically different way 
\cite{BPcoordinatesRolcrossterm,RolCrossTermNix}.
Within the framework of a given model, this new data may allow 
an estimate of $\Delta\tau$, 
(and also other parameters described in the references)
as a function of beam energy, using a variable that 
has different dependence on $\Delta\tau$ than does the more commonly studied 
quantity $R^2_{\mathrm{out}}-R^{2}_{\mathrm{side}}$.  

    One other observation can be made because the $R^{2}_{ol,0}$ values
in Fig.~\ref{F26} are measured in the LCMS frame.  As mentioned above,
non-zero values of $R^{2}_{ol,0}$ suggest boost-invariance may be broken.
The higher absolute values of $R^{2}_{ol,0}$ at lower $\sqrt{s_{NN}}$ may thus 
reflect that the assumption of boost-invariance becomes less valid at lower energies.

\begin{figure}
\includegraphics[width=3.4in]{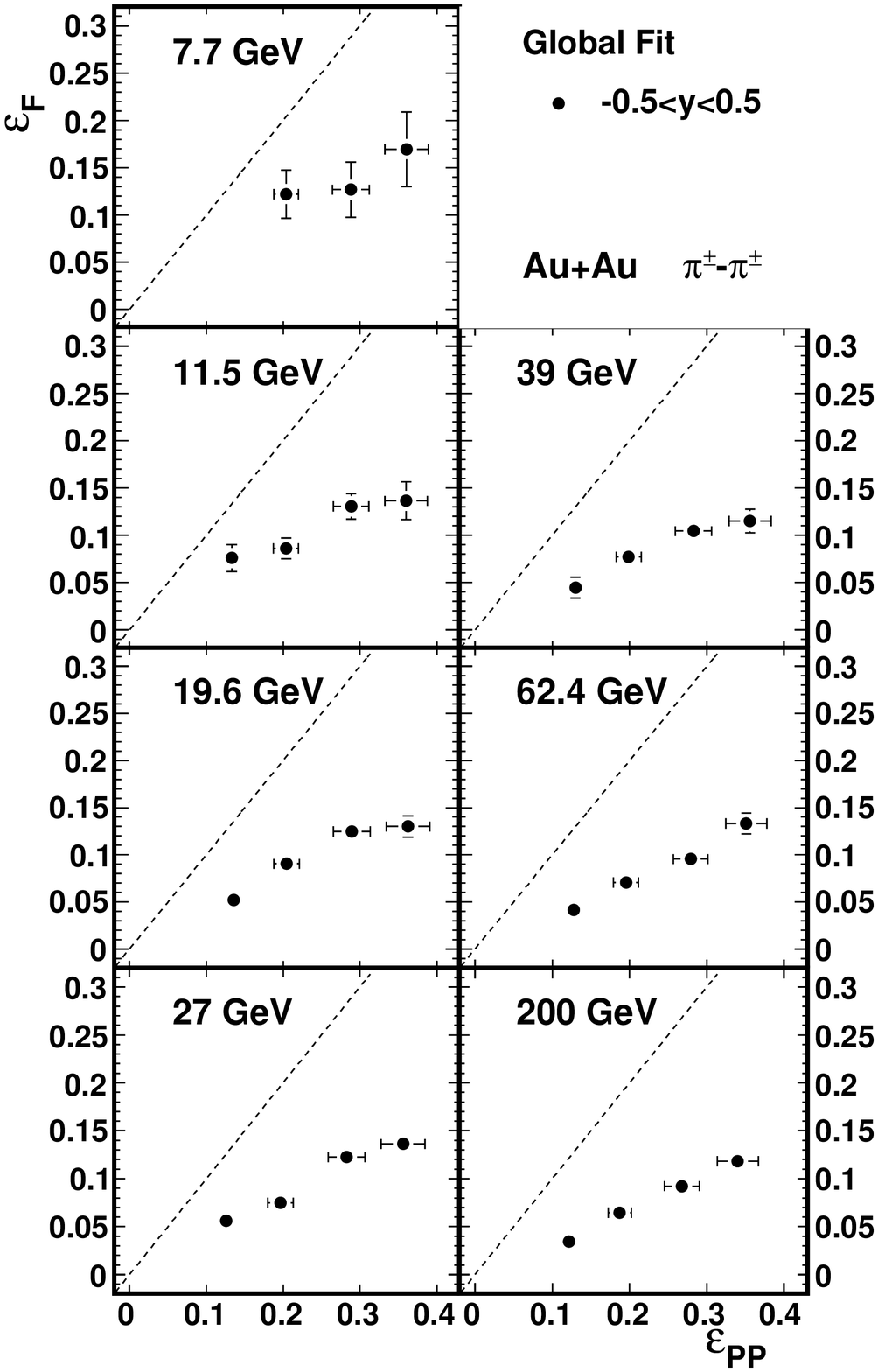}
\caption{\label{F24}  The eccentricity of the collisions at
kinetic freez-out, $\varepsilon_{F}$, as a function of initial eccentricity
relative to the participant plane, $\varepsilon_{PP}$, at mid-rapidity.  All results are for $\langle k_{T}\rangle\approx$ 0.31 GeV/$c$.  Error bars include only statistical uncertainties.  The line has a slope of one indicating no change in shape.  Points further below the line evolve more to a round shape.}
\end{figure}

\subsubsection{Kinetic freeze-out eccentricity}\label{S5b3}

  Once the Fourier coefficients are extracted the eccentricity, defined as
  
\begin{equation}\label{E10}
\varepsilon_{F} =
\frac{\sigma '^{2}_{y}-\sigma '^{2}_{x}}{\sigma '^{2}_{y}+\sigma '^{2}_{x}}\approx2\frac{R^{2}_{s,2}}{R^{2}_{s,0}}
\end{equation}
can be simply computed 
\cite{LisaRetiere}.  The variances
$\sigma '_{y}$ and $\sigma '_{x}$ correspond to the widths of the collision fireball
at kinetic freeze-out in the out-of-plane and in-plane directions, respectively.
This definition allows negative eccentricities if 
$\sigma '_{y}<\sigma '_{x}$ 
which would indicate the system expanded enough to become in-plane extended.
Whether or not that happens is related to the collision dynamics and equation of
state as described in Sec.~\ref{S2}. The ratio $R^{2}_{s,2}/R^{2}_{s,0}$ is 
used to estimate $\varepsilon_{F}$ because $R_{\mathrm{side}}$ is less affected by 
flow, and hence it carries primarily geometric information \cite{LisaRetiere}.

  Figure~\ref{F24} shows the eccentricities at kinetic freeze-out, 
$\varepsilon_{F}$, defined in Eq.~\ref{E10}, for all centralities and energies.  
They are plotted against the initial eccentricity relative to the
participant plane obtained from the Glauber model~\cite{StarBesFlowEccentricity}, 
defined as

\begin{equation}
\varepsilon_{PP}=\frac{\sqrt{(\sigma^{2}_{y}-\sigma^{2}_{x})^{2}+4\sigma^{2}_{xy}}}{{\sigma^{2}_{x}+\sigma^{2}_{y}}}.
\end{equation}
The variances $\sigma^{2}_{x}=\{x^{2}\}-\{x\}^{2}$ and $\sigma^{2}_{y}=\{y^{2}\}-\{y\}^{2}$ 
gauge the widths of the distributions of participant nucleons in and out of
the reaction plane direction, respectively.  The symbol $\{\mathellipsis\}$ 
denotes averaging of participant nucleons, with positions $x$ and $y$, in each event. 
The covariance $\sigma_{xy}=\{xy\}-\{x\}\{y\}$ accounts for event-by-event
fluctuations in the distribution of participant nucleons.  
The line has a slope 
of one ($\varepsilon_{F}=\varepsilon_{PP}$), so points further below the line 
have evolved more toward a round shape ($\varepsilon_{F}=0$).  These results 
demonstrate that, at all energies studied, the freeze-out shape 
remains an out-of-plane extended ellipse ($\varepsilon_{F}>0$).  In no case does 
extended lifetime or stronger flow result in the shape becoming in-plane 
extended ($\varepsilon_{F}<0$).  However, there is always some evolution toward 
a more round shape, as expected, and there tends to be slightly more evolution 
for the higher energies.  The same observations apply at forward and backward
rapidity because of the similar trends observed for the ratio 
$R^{2}_{s,2}/R^{2}_{s,0}$ ($=\varepsilon_{F}/2$).

\begin{figure*}
\includegraphics[width=6.0in]{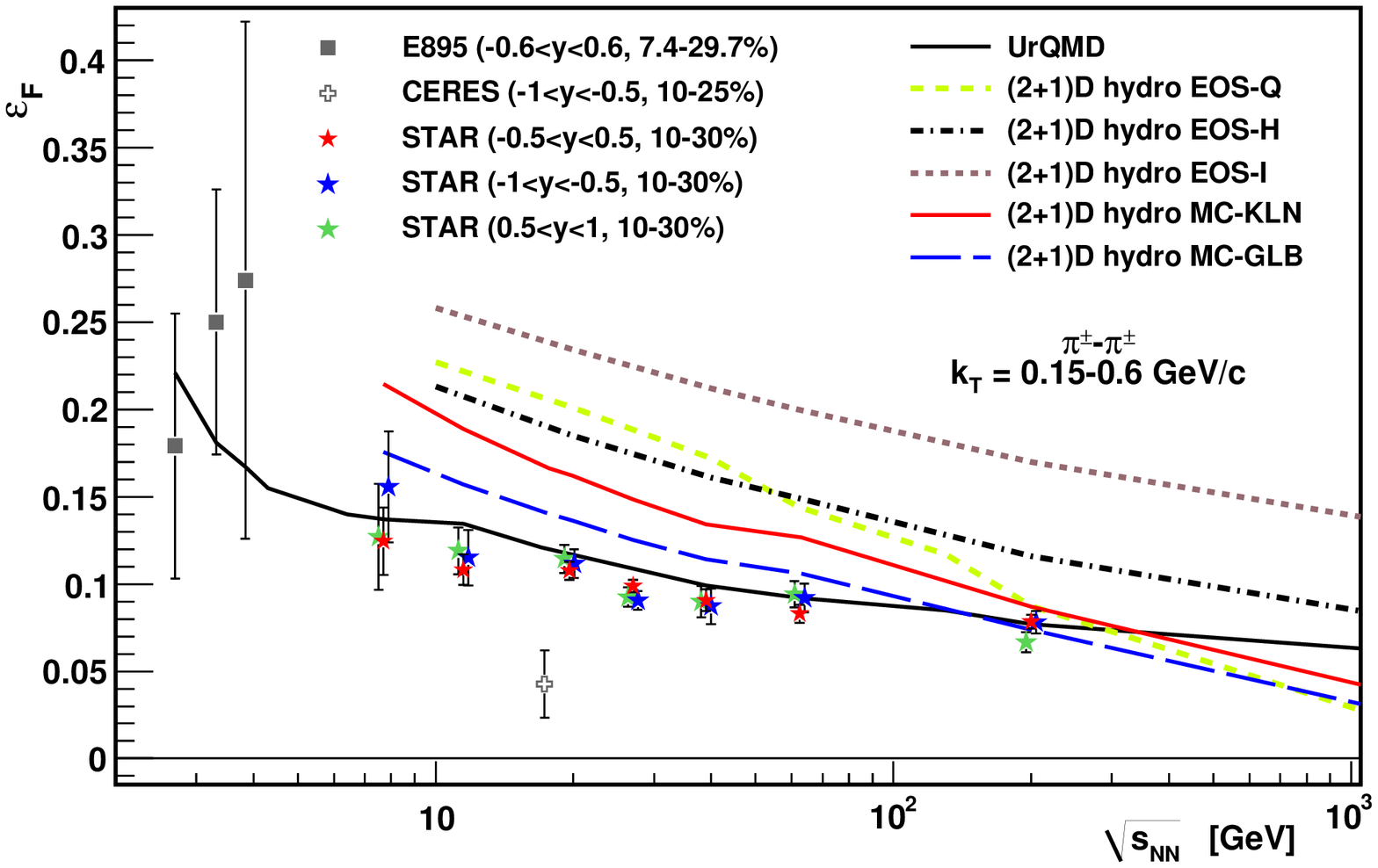} 
\caption{\label{F25}  (Color online) The dependence of the kinetic freeze-out eccentricity of pions on collision energy in mid-central Au+Au collisions (E895, STAR) and Pb+Au collisions (CERES) for three rapidity regions and with $\langle k_{T}\rangle\approx$ 0.31 GeV/$c$.  For clarity, the points for forward and backward rapidity from STAR are offset slightly.  Error bars include only statistical uncertainties.    Several (2+1)D hydrodynamical models and UrQMD calculations are shown.  Model centralities correspond to the data.  The trend is consistent with a monotonic decrease in eccentricity with beam energy.}
\end{figure*}

  The excitation function for the freeze-out shape is presented in Fig.~\ref{F25}.
The new STAR results for three rapidities are compared to earlier measurements
from other experiments and to several models.  The results use the global fit
method and are for mid-peripheral ($10$-$30\%$) collisions where 
the initial anisotropic shape is large but there is still significant overlap
of the nuclei.  The larger differences between in-plane and out-of-plane
pressure gradients in these collisions and larger initial spatial anisotropy 
could admit more varied results in the change in shape, if that where to happen
at different energies.  The new STAR results exhibit a monotonic decrease in the 
freeze-out eccentricity with increasing beam energy for all three rapidity regions.  

The freeze-out eccentricity values from CERES and STAR at similar energy and 
centrality are not consistent.
  There are some differences in analyses from these different
experiments such as correction for event plane resolution, fitting in one 
$k_{T}$ bin versus averaging several smaller $k_{T}$ bins, and centrality 
ranges.  These could potentially be important and were studied.  The CERES
point at 17.3 GeV suggested a possible minimum in the historical data.  The
new STAR results at 11.5 and 19.6 GeV at mid-rapidity were significantly 
higher suggesting a monotonic decrease in the freeze-out shape.  To check 
that the difference was not due to the different rapidity windows the STAR 
analysis was extended to include the same rapidity region as CERES, $0.5<|y|<1$.
The forward and backward rapidity results remained consistent with the 
mid-rapidity measurement.  
The CERES point for $10$-$25\%$ centrality is consistent with the (smaller)
eccentricities for the $0$-$5\%$ and $5$-$10\%$ centrality ranges in STAR 
results at 19.6 GeV, so it seems rather unlikely that large enough differences 
in centrality definitions could occur to cause such a large difference in the 
eccentricities for STAR and CERES.  Event, track, and
pair selection quantities have rather little effect on the results.  Another
difference is the range of $k_{T}$ values included in the fits.  In the
CERES and earlier STAR result \cite{PRLstarAziAuAu200}, the 
azimuthal analysis was done in narrow
$k_{T}$ bins and the $\varepsilon_{F}$ values averaged.  This was problematic
at the lowest energies due to lower statistics when the analysis was
additionally differential in $k_{T}$.  Using a single, wide $k_{T}$ bin
biases the results slightly toward smaller $\varepsilon_{F}$ values, as
discussed in Sec.\ref{S3d}.  Therefore, to be consistent,
the same (wide $k_{T}$ bin) method is used for all the STAR points.  
The CERES results used a weighted average of results in narrow $k_{T}$ bins 
which should be equivalent to using a single, wide $k_{T}$ bin.  It seems 
unlikely that this is the cause of the discrepency.
The E895 
correction algorithm was used in the CERES and E895
cases to correct for the event plane resolution while in the STAR case
the histograms were corrected or the fit function smeared in the global fit
case.  The difference in the results is rather tiny for these different methods
and also cannot explain the difference.  

  As discussed in Sec.~\ref{S2}, non-monotonic behavior in the excitation
function would have strongly suggested interesting changes in the equation
of state.  The observed monotonic decrease excludes the scenario described in
reference \cite{ShapeAnalysis} and is
consistent with increased lifetime and/or pressure gradients at higher energy.
The energy dependence of $R_{\mathrm{long}}$ from the non-azimuthal analysis,
and the lifetimes shown in Fig.~\ref{F9c}, suggest
also that the system is longer-lived at higher energy.  Still, these results
will allow to probe equation of state effects by comparing to various models.

  The currently available model predictions 
\cite{KolbHeinzHydro,ShenEpsFvsRootShydro,ShapeAnalysis} for the energy 
dependence of the 
freeze-out eccentricity are also shown in Fig.~\ref{F25}.  All models predict a 
monotonic decrease in the freeze-out shape at higher energies similar to the data.
The older (2+1)D, ideal hydrodynamical 
models \cite{KolbHeinzHydro}, labeled EOS-H, EOS-I, and EOS-Q, all overpredict 
the  data.  As was noted in \cite{ShenEpsFvsRootShydro}, in comparison to the 
historical data, the model with a first order phase transition, EOS-Q, gets 
close to the 200 GeV point.  The predictions of the freeze-out shape are 
sensitive to the equation of state used in the hydrodynamic models.  This is 
clear by comparing the curves for EOS-I (ideal, massless quark gluon gas) and 
EOS-H (hadronic gas).  For EOS-Q, the slope changes, following EOS-H at
low energies, but dropping more rapidly at higher energies.  This is attributed
to passage through a mixed-phase regime which extends the lifetime allowing the
system to evolve to a more round state at higher energies \cite{ShapeAnalysis}.

  The two more recent (2+1)D predictions, from the VISH2+1 model, get closer 
to the data.  MC-KLN and MC-GLB correspond to different initial conditions and
are more realistic than the earlier results as they allow to 
incorporate viscous effects~\cite{ShenEpsFvsRootShydro}.  
MC-GLB uses a specific shear viscosity of 
$\eta/s=0.08$ with Glauber initial conditions.  
The MC-KLN model has a much larger specific shear viscosity, $\eta/s=0.2$, and 
the initial shape is 
derived from the initial gluon density distribution in the transverse plane 
(which is converted to an entropy and finally energy density profile).
Both models incorporate an equation of state 
based on lattice QCD, named s95p-PCE~\cite{HydroEosRefA,HydroEosRefB}. 
Initial parameters in the models were calibrated using measured multiplicity
distributions (and extrapolations to lower energies) and to describe
$p_{T}$-spectra and $v_{2}$ measurements for 200 GeV Au+Au collisions at
RHIC. 
The two cases were found to yield similar lifetimes, but in the
MC-KLN case the initial eccentricities are larger (more out-of-plane extended).
The MC-KLN model achieves a less round shape simply because it starts with
larger initial eccentricity \cite{ShenEpsFvsRootShydro}.  The excitation
function for freeze-out eccentricities has the potential to resolve ambiguities
between models with different initial conditions and values of $\eta/s$.  In
particular, the two sets of initial conditions and $\eta/s$ used here yield identical 
$v_{2}$, but very different $\varepsilon_{F}$.  So the results in 
Fig.~\ref{F25} provide tighter constraints on these models.

  The goal of \cite{ShenEpsFvsRootShydro} was to map systematic trends in 
observables with the two models, not to explain the data precisely.  In fact, 
the applicability of these models is known to be problematic at lower energies 
both because they assume boost-invariance, which is broken at lower energy, and
because the hadronic phase is expected to become more important at lower 
collision energy.  
A more realistic calculation requires (3+1)D viscous hydrodynamics.
Nevertheless,
the new calculations are able to match more closely the experimental results.
Of the hydrodynamical models, MC-GLB is closest to the data although it still 
overpredicts the freeze-out eccentricity and the slope appears too steep.  
One relevant observation from \cite{ShenEpsFvsRootShydro} is that in these
models the decrease in the eccentricity with energy appears to be due mainly to 
an extended lifetime rather than larger anisotropy of pressure gradients.
As discussed at the end of Sec.~\ref{S5a}, the lifetime extracted from $R_{\mathrm{long}}$ values
also suggest an increase in the total lifetime.  However, the data cannot allow 
one to determine whether the decrease in eccentricity is due solely to increased 
lifetime or whether the pressure gradients may also play a significant role.

  The prediction of the Boltzmann transport model, UrQMD (v2.3) \cite{urqmdnewver}, 
matches most closely the freeze-out shape at all energies \cite{ShapeAnalysis}.  
UrQMD follows the 
trajectories and interactions of all hadronic particles throughout the
collision, 
so it does not require assumptions about how freeze-out occurs.  The model is 3D 
and does not require boost-invariance, therefore it is equally applicable at all the 
studied energies.  This may be, at least partially, why the predictions from UrQMD 
more closely match the energy dependence of the data compared to the hydrodynamic predictions.  
While it does not explicitly contain a deconfined state, it does incorporate color 
degrees of freedom through inclusion of the creation of color strings and their 
subsequent decay back into hadrons.

  For the azimuthally integrated results, UrQMD does rather well at predicting
the observed dependence of HBT radii on $\langle k_{T}\rangle$ and centrality 
\cite{urqmdPRC73_064908_2006,urqmdJPG34_537_2007}.  As discussed earlier, 
inclusion of a mean field acting between pre-formed hadrons (color string fragments) 
predicts $R_{\mathrm{out}}/R_{\mathrm{side}}$ ratios similar to the observed 
values and leads naturally to a minimum in the volume similar to that which is 
observed experimentally \cite{urqmdPRB659_525_2008,urqmdVolumeRef}.  Such
a repulsive potential between the string fragments would mimic somewhat an
increase in pressure gradients at early stages \cite{urqmdPRB659_525_2008} 
similar to the hydrodynamics cases with an equation of state that includes a 
phase transition.  The UrQMD predictions for the eccentricities at kinetic 
freeze-out in Fig.~\ref{F25} were made with UrQMD in cascade mode and so do not 
incorporate this potential between string fragments.  

  It should be noted that none of the models 
predict
all observables simultaneously.  The UrQMD model, while it matches the 
freeze-out shapes well, matches the momentum space observables less well.  
And the hydrodynamic models, while they are able to describe the momentum space
$p_{T}$ spectra and $v_{2}$ results, do less well at predicting the 
eccentricity and trends observed in HBT analyses.  The availability of 
these new experimental results provide an important opportunity to further 
constrain models.  

\section{Conclusions}\label{S6}

  The two-pion HBT analyses that have been presented provide key measurements
in the search for the onset of a first-order phase transition in Au+Au
collisions as the collision energy is lowered.  The Beam Energy
Scan program has allowed HBT measurements to be carried out across a wide
range of energies with a single detector and identical analysis techniques.  In
addition to standard azimuthally integrated measurements, we have performed
comprehensive, high precision, azimuthally sensitive femtoscopic measurements of
like-sign pions.  In order to obtain the most reliable estimates of the
eccentricity of the collisions at kinetic freeze-out, a new global fit method 
has been developed.

  A wide variety of HBT measurements have been performed and the comparison of
results at different energies is greatly improved.  In the azimuthally
integrated case, the beam energy dependence of the radii generally agree with 
results from other experiments, but show a much smoother trend than the earlier 
data which were extracted from a variety of experiments with variations in 
analysis techniques.  The current analyses additionally contribute 
data in previously unexplored regions of collision energy.  The transverse mass
dependence is also consistent with earlier observations and allows one to 
conclude that all $k_{T}$ and centrality bins exhibit similar trends as a
function of collision energy.


  The energy dependence of the volume of the homogeneity regions is consistent
with a constant mean free path at freeze-out, as is the very flat energy 
dependence of $R_{\mathrm{out}}$.  This scenario also explains the
common dependence of $R_{\mathrm{side}}$ and $R_{\mathrm{long}}$ on the
cube root of the multiplicity that is observed at higher energy.  For
7.7 and 11.5 GeV, $R_{\mathrm{side}}$ appears to deviate slightly from the 
trend at the higher energies.  Two physical changes that may potentially be
related to this are the effects of strangeness enhancement (not included in the 
argument for a constant mean free path at freeze-out ) and the rapid
increase in the strength of $v_{2}$ that levels off around 7.7 to 11.5 GeV.  
Both of these physical changes occur in the vicinity of the minimum.  A 
systematic study with a single detector at slightly lower energies would be 
needed to help disentangle the different effects.  

   The UrQMD model provides an alternative explanation for the minimum in the
volume measurement in terms of a change from a hadronic to a partonic state.  
Including interactions between color string fragments early in the collision, it
not only can explain the minimum in the volume, but is also able to find 
$R_{\mathrm{out}}/R_{\mathrm{side}}$ values close to unity as observed from AGS
through RHIC energies and improves the agreement between UrQMD and other 
observables at the same time.  It is interesting that such
an interaction potential may somewhat mimic an increase in the pressure
gradients, which may correlate with the observation that $v_{2}$ increases
rapidly with $\sqrt{s_{NN}}$ in this region also.

  The lifetime of the collision evolution was extracted using the 
$\langle m_{T}\rangle$ dependence of $R_{\mathrm{long}}$.  Subject to certain
assumptions, the lifetime increases by a factor 1.7 from AGS to 200 GeV
collisions measured at STAR.  The lifetime increases by about 1.4 times more 
between RHIC and the LHC.

  A new global fit method was developed and studied in relation to the HHLW
fit method.  
For most centralities, this method is found to yield more reliable results
in cases of low statistics and poor event plane resolution, although it has
problems in the most central bin related to different parameterizations.
Additionally, it avoids problems related to correlated errors.  This global fit method has 
allowed the extraction of the most reliable results at the lowest energies 
studied.

  The Fourier coefficients measured away from mid-rapidity allow one to extract 
the energy dependence of the $R^{2}_{ol,0}$ cross term.  This previously unavailable 
observable exhibits a monotonic decrease as a function of beam energy.  This observable 
has been connected to the duration of 
particle emission in a way that is different than the more commonly studied quantities 
$R^{2}_{\mathrm{out}}-R^{2}_{\mathrm{side}}$ or $R_{\mathrm{out}}/R_{\mathrm{side}}$.  
This measurement may provide constraints for models that relate the radii
and physical quantities with different sets of assumptions.

  The azimuthally differential results show that, for all energies, the
evolution of the collision eccentricity leaves the system still out-of-plane
extended at freeze-out.  In mid-central ($10$-$30\%$) collisions, the freeze-out
eccentricity shows a monotonic decrease with beam energy consistent with 
expectations of increased flow and/or increased lifetime at higher energies.
This is supported by the azimuthally integrated results which suggest longer
lifetimes at higher energies.  The results are consistent qualitatively with the
monotonic decrease suggested by all model predictions, but is most consistent 
quantitatively with UrQMD.  While 
the hydrodynamic models can match momentum space observables 
($p_{T}$ spectra, $v_{2}$) well, they do less well at predicting the HBT 
results.  At the same time, while the UrQMD model does better at predicting 
the HBT results, like the freeze-out shape, it does less well at predicting 
the momentum space observables.  The freeze-out eccentricity excitation 
function provides new, additional information that will help to constrain 
future model investigations.

\appendix*
\section{Non-Gaussian effects on azimuthal HBT analyses}\label{S4h}

  In azimuthally integrated HBT analyses, the cross terms ($R_{os}$, $R_{ol}$, 
$R_{sl}$) vanish at mid-rapidity.  In this case, the sign of the components
of the relative momentum vector, $\vec{q}$, are arbitrary.  The three dimensional
$\vec{q}$-space distributions (numerator, denominator, and Coulomb weighted distributions) 
may be folded, so that $q_{\mathrm{long}}$ and $q_{\mathrm{side}}$ are always 
positive, for instance, to increase statistics in each
($q_{\mathrm{out}}$,$q_{\mathrm{side}}$,$q_{\mathrm{long}}$) bin.  In 
azimuthally differential analyses, 
however, the relative signs of components are important 
in order to extract non-zero cross terms \cite{RWellsThesisQfolding,PRCstarAuAu200}.  
At mid-rapidity, the relative sign of $q_{\mathrm{out}}$ and $q_{\mathrm{side}}$ must thus be
maintained to extract values of $R^{2}_{os}$.  Away from mid-rapidity, the 
$R^{2}_{ol}$ cross term is also non-zero and $q_{\mathrm{long}}$ must be allowed to have 
both positive or negative values.  This way the relative sign of $q_{\mathrm{out}}$ and 
$q_{\mathrm{long}}$ is maintained and the corresponding cross term can be extracted.

  If the ``$q$-folding'' procedure is performed 
and the cross terms are included as fit parameters, the fit parameters become
strongly correlated and the values of the extracted radii change.  The size
of this effect varies randomly from one azimuthal bin to the next, causing large
variations in the extracted oscillations of the radii.  This behavior is
related to the non-Gaussianess of the correlation function.  Due to the
necessity of using finite bins in $k_{T}$ and centrality, which are described 
by a range of radii, the radii extracted from these correlation functions are 
some average value.  If too much $q$-folding is performed the signs
of the relative momentum components are lost.  In cases where the cross terms
associated with these relative momentum components are non-zero, the covariance of fit 
parameters that appears allows deviations from the average values and the
results become unreliable.

  This is an important consideration for any HBT analysis performed away from
mid-rapidity, or relative to the first order reaction plane, where measurement
of cross terms is important.  In this analysis, no folding of $\vec{q}$-space
is performed and so any possible effects of this phenomena are eliminated.

We thank the RHIC Operations Group and RCF at BNL, the NERSC Center at LBNL, the KISTI Center in Korea and the Open Science Grid consortium for providing resources and support. This work was supported in part by the Offices of NP and HEP within the U.S. DOE Office of Science, the U.S. NSF, CNRS/IN2P3, FAPESP CNPq of Brazil, Ministry of Ed. and Sci. of the Russian Federation, NNSFC, CAS, MoST and MoE of China, the Korean Research Foundation, GA and MSMT of the Czech Republic, FIAS of Germany, DAE, DST, and CSIR of India, National Science Centre of Poland, National Research Foundation (NRF-2012004024), Ministry of Sci., Ed. and Sports of the Rep. of Croatia, and RosAtom of Russia.


\bibliography{References} 

%

\providecommand{\noopsort}[1]{}\providecommand{\singleletter}[1]{#1}%


\end{document}